\documentclass[12pt]{article}

\usepackage{ graphicx, amsmath, amsthm, color, psfrag, amsfonts, natbib,
  url, amssymb, bm, thmtools, mathtools }
  
\usepackage[affil-it]{authblk}

\usepackage{endnotes}

\usepackage[utf8]{inputenc}

\let\footnote=\endnote

\newcommand{\given}{\, | \,}

\newcommand{\x}[1]{$O \! \left( n^{ - \frac{ 1 }{ 2 } } \right)$}

\newcounter{remark}
\newcommand{\remark}{%
  \stepcounter{remark}%
  \theremark}

\declaretheorem[ style = definition, numbered = yes ]{definition}

\declaretheorem[ style = theorem, numbered = yes ]{theorem}
\declaretheorem[ style = lemma, numbered = yes ]{lemma}

\begin{document}

\title{\mbox{} \vspace*{-1.0in} \\ \textbf{The Practical Scope \\ of the \textit{Central Limit Theorem}}}

\author{\Large{\textsf{David Draper}}\footnote{\textit{Address for
correspondence}: David Draper, Department of Statistics, Baskin School of Engineering, University of California, 1156 High Street, Santa Cruz CA 95064 USA; email \texttt{draper@ucsc.edu} . \\ \hspace*{0.15in} $^\dagger$University of California, Santa Cruz, email: \texttt{eguo1@ucsc.edu} .} \ and \textsf{Erdong Guo$^\dagger$} \vspace*{-0.1in}}

\affil{\large University of California, Santa Cruz \vspace*{-0.25in}}

\date{23 Nov 2021}

\maketitle

\vspace*{-0.5in}

\begin{abstract}

\noindent
The \textit{Central Limit Theorem (CLT)} is at the heart of a great deal of applied problem-solving in statistics and data science, but the theorem is silent on an important implementation issue: \textit{how much data do you need for the CLT to give accurate answers to practical questions?} Here we examine several approaches to addressing this issue --- along the way reviewing the history of this problem over the last 290 years --- and we illustrate the calculations with case studies from finite-population sampling and gambling. A variety of surprises emerge.

\bigskip

\noindent
\textbf{Keywords:} Jacob Bernoulli, continuity correction, Cornish-Fisher expansion, Abraham de Moivre, Edgeworth expansion, excess kurtosis, false-discovery errors, Pierre Simon de Laplace, lattice distribution, replication crisis, skewness.

\end{abstract}

\section{Introduction} \label{s:introduction-1}

The \textit{Central Limit Theorem} (CLT) has been a bedrock of probability theory since Abraham \cite{de-moivre-1733} was the first to derive the Normal approximation to the Binomial distribution in the 1730s and Pierre Simon de \citet{laplace-1810} proved the first CLT in 1810, for general discrete distributions (we summarize the main history of the CLT in Section \ref{s:history-1} below). \cite{laplace-1778} was also among the first\footnote{The first person to address this question was John \cite{arbuthnot-1712}, who (unlike Laplace) used a non-parametric procedure (the sign test) to reject the same hypothesis. Starting intermittently in 1592 and then regularly in 1603, every Thursday the government in the city of London printed \textit{Bills of Mortality} (one for each parish), which recorded burials and christenings happening over the previous week. John \cite{graunt-1662} published analyses of this data set in the mid-1660s; he seems to have been the first person to notice that the male birth rate was consistently (year upon year) higher than the female rate. Arbuthnot continued Graunt's work by looking at the same source of data from 1629 to 1710, noting that for 82 consecutive years more males were born than females. This gave him what we would now call a sign-test $p$-value of $2^{ -82 } \doteq 10^{ -25 }$.} to have used the CLT in what we would now call parametric statistical inference, to reject the hypothesis that the probability of a male human birth is 0.5 (in most countries today, it's around 0.512, as it was in Laplace's time).

In its simplest form, the CLT is a theorem about probability distributions for real-valued uncertain quantities, modeled as realizations of a random variable\footnote{We operate here implicitly within the standard Kolmogorov foundational framework with its probability space $( \Omega, \mathcal{ F }, P )$, in which a real-valued random variable $Y$ is a $P$--measurable function from $\Omega$ to $\mathbb{ R }$. In the red-or-black roulette example of Section \ref{s:roulette}, for instance, in which $Y$ represents the net gain for the gambler (not the house) after a single 1--monetary-unit (MU) play on red (say), one possible specification is as follows: $\Omega = \{ \omega_1, \omega_2, \omega_3 \} = \{ \textrm{black, green, red} \}$, $\mathcal{ F }$ is the power set $2^\Omega$ of $\Omega$, $P$ assigns probabilities $\left( \frac{ 18 }{ 38 }, \frac{ 2 }{ 38 }, \frac{ 18 }{ 38 }\right)$ to the atoms of $\Omega$ listed in alphabetical order, and $Y ( \omega_i ) = -1$ MU for black, $-1$ MU for green, and $+1$ MU for red.} $Y$; we focus here on this setting, and throughout the paper (for concision) we use the abbreviation \textit{random variable} for the phrase \textit{real-valued random variable}. Distributions in this context may be characterized in a number of ways, of which the following three appear to be the most useful in practical data science: \vspace*{-0.1in}
\begin{itemize}

\item

the \textit{cumulative distribution function} (CDF) $F_Y ( y ) = P ( Y \le y )$, which always exists for all $y \in \mathbb{ R }$; 

\item

the \textit{probability mass function} (PMF) $f_Y ( y ) = P ( Y = y )$ of a discrete random variable, which always exists if the support of $Y$ contains at most countably many possible values, or the \textit{probability density function} (PDF) $f_Y ( y ) = \frac{ d }{ d y } F_Y ( y )$ of a continuous random variable, if $F_Y$ is an absolutely continuous function (other possibilities do arise with some frequency in practical data science;
for example, when modeling daily rainfall amounts at a particular location it's natural to employ a \textit{mixed} distribution with a PMF point-mass at 0 mm and a PDF for the positive values).

\item

the \textit{inverse CDF} or \textit{quantile function} $F_Y^{ -1 } ( p )$, defined\footnote{Throughout the paper, the symbol $\triangleq$ means \textit{is defined to be}.} in general for $0 < p < 1$ as 
\begin{equation} \label{e:quantile-0}
y_p \triangleq F_Y^{ -1 } ( p ) \triangleq \inf_{ y \in \mathbb{ R } } \{ y \! : F_Y ( y ) \ge p \} \, .
\end{equation}
This definition is needed for (discrete and continuous) settings in which the CDF has one or more flat spots (constant regions); in the absence of such flat spots, the quantile function is defined uniquely by the simple relation $y_p = F_Y^{ -1 } ( p ) \textrm{ iff } F_Y ( y_p ) = p$.

\end{itemize}
Other ways in which the probability behavior of $Y$ may be quantified include its \textit{characteristic function} (CF), which plays in important behind-the-scenes (but not in-the-foreground, except in Section \ref{s:history-1}) role in the work presented here\footnote{For example, one of the simplest proofs of the CLT relies on the \textit{L\'evy Continuity Theorem}, which relates convergence in distribution of the $Z_n$ in \textbf{Theorem \ref{t:clt-on-cdf-scale-1}} (Section \ref{s:introduction-1}) to pointwise convergence of the CFs of the $Z_n$; see, e.g., \cite{williams-1991}.}.

\begin{quote}

\textbf{Remark \remark.} It's possible to create probability distributions that can accurately be described as pathological from the viewpoint of applied data science; as an example (see, e.g., \cite{dovgoshey-2006}), the Cantor function (when extended to include the values 0 and 1 on $( - \infty, 0 )$ and $( 1, \infty )$, respectively) satisfies all properties of a CDF with support on the unit interval and is everywhere continuous but has 0 derivative almost everywhere on its support. We wish to exclude all such pathologies in this paper. To be precise about our exclusion, consider \textit{Lebesgue's Decomposition Theorem} (e.g., \cite{halmos-1974}) when applied to probability measures on the real line, which states that any CDF $F_X ( x )$ of a random variable $X$ can be expressed as the mixture
\begin{equation} \label{e:lebesgue-decomposition-1}
F_X ( x ) = \pi_1 \, F_X^D ( x ) \, + \, \pi_2 \, F_X^C ( x ) \, + \, \pi_3 \, F_X^S ( x ) \, ,
\end{equation} 
in which $( F_X^D, \, F_X^C, \, F_X^S )$ are CDFs of (discrete, (absolutely) continuous, singular\footnote{Note that a probability distribution $P ( \, \cdot \, )$ on $\mathbb{ R }$ is \textit{singular} with respect to $( \lambda \triangleq$ Lebesgue measure on $\mathbb{ R } )$ iff $P ( \, \cdot \, )$ concentrates all of its probability on a set of $\lambda$-measure 0.}) random variables (respectively) and where $0 \le \pi_i \le 1$ (for $i = 1, 2, 3$) with $\sum_{ i = 1 }^3 \pi_i$. Weirdness can ensue whenever $\pi_3 > 0$. In this paper $\pi_3 = 0$, i.e., we exclude all CDFs with singular components; such CDFs \textit{never} arise in practical data science.

\end{quote}

The CLT is most frequently stated on the CDF scale, but CLTs also exist on the other two scales mentioned above (PMF/PDF and inverse CDF); we examine all three in what follows, beginning with CDFs. For subsequent reference, as usual let $\Phi ( z )$ and $\phi ( z )$ be the CDF and PDF of the standard Normal distribution, respectively. \vspace*{-0.35in}

\begin{quote}

\begin{theorem} \label{t:clt-on-cdf-scale-1}

\textbf{(CLT on the CDF scale).} With $n$ as a positive integer, let $\{ Y_i, \, i = 1, \dots, n \}$ be IID random variables with mean $\mu$ and variance $\sigma^2 > 0$, both of which are assumed to exist and to be finite, and set $\bar{ Y }_n = \frac{ 1 }{ n } \sum_{ i = 1 }^n Y_i$. Then 
\begin{equation} \label{e:clt-1}
\textrm{as } n \rightarrow \infty \, , \ \ \ \frac{ \left( \bar{ Y }_n - \mu \right) }{ \sigma / \sqrt{ n } } \stackrel{ D }{ \rightarrow } N ( 0, 1 ) \, .
\end{equation}
Another way to express this fact is to define, for $y \in \mathbb{ R }$, 
\begin{equation} \label{e:edgeworth-1}
Z_n \triangleq \frac{ \bar{ Y }_n - \mu }{ \sigma / \sqrt{ n } } \, , \ \ \ z \triangleq \frac{ y - \mu }{ \sigma / \sqrt{ n } } \, , \ \ \ \textrm{and} \ \ \ F_{ Z_n } ( z ) \triangleq P ( Z_n \le z ) \, ;
\end{equation}
equation (\ref{e:clt-1}) is then equivalent to the statement
\begin{equation} \label{e:clt-2}
\textrm{for all } z \in \mathbb{ R }, \ \ \ \left| F_{ Z_n } ( z ) - \Phi ( z ) \right| \rightarrow 0 \ \ \ \textrm{as} \ \ \  n \rightarrow \infty \, .
\end{equation}
Moreover, by a result due to Georg \cite{polya-1920}, the convergence in (\ref{e:clt-2}) is \textit{\textbf{uniform}} in $z$:
\begin{equation} \label{e:clt-3}
\lim_{ n \rightarrow \infty} \sup_{ z \in \mathbb{ R } } \left| F_{ Z_n } ( z ) - \Phi ( z ) \right| \rightarrow 0 \, .
\end{equation}

\end{theorem}

Informally \textbf{Theorem \ref{t:clt-on-cdf-scale-1}} says that, for large $n$, $\bar{ Y }_n$ is approximately Normally distributed with mean $\mu$ and variance $\frac{ \sigma^2 }{ n }$.

\end{quote} 

Note that the single necessary and sufficient condition for \textbf{Theorem \ref{t:clt-on-cdf-scale-1}} to hold, above and beyond the all-important IID assumption, is $0 < V ( Y_i ) \triangleq \sigma^2 < \infty$, which gives the \textbf{Theorem} remarkably wide scope as a \textit{limiting} result. However, for applied statisticians and data scientists, a natural implementation question immediately arises: 

\begin{quote}

$\mathbb{ Q }$: \textit{When is $n$ large enough so that the approximation of $F_{ Z_n } ( z )$ by $\Phi ( z )$, \\ \hspace*{0.2in} for interesting values of $z$, is good?} 

\end{quote}

The plan of the rest of the paper is as follows. In Section \ref{s:history-1} we offer a short history of the CLT. Section \ref{s:accuracy-quantification} reviews and discusses the literature on ways to quantify the distance between $F_{ Z_n } ( \cdot )$ and $\Phi ( \cdot )$ in \textbf{Theorem \ref{t:clt-on-cdf-scale-1}}. In Section \ref{s:edgeworth-cornish-fisher} we 

\begin{itemize}

\item[(a)] 

consider skewness and excess kurtosis of the underlying distribution $F_Y ( \cdot )$ as quantitative indicators of non-Normality and 

\item[(b)] 

examine the value of Edgeworth and Cornish-Fisher asymptotic expansions out to orders $n^{ - \frac{ 1 }{ 2 } }$ and $n^{ - 1 }$ in answering question $\mathbb{ Q }$ above. 

\end{itemize}
We address continuous, discrete, and mixed $Y_i$ in Section \ref{s:edgeworth-cornish-fisher}, and we study the CLT on all three scales mentioned above (CDF, PDF/PMF, and quantile). In Section \ref{s:case-studies} we present detailed calculations in two case studies, drawn from the worlds of IID random sampling from finite populations and gambling. Section \ref{s:discussion} offers a brief discussion and some concluding comments.

%; an Appendix rounds out the paper with details on CLT calculations on the CDF scale with lattice distributions.

% edit this if we do something different in the appendix or if we have no appendix at all

By way of additional reading on the topic of this paper, many authors have made significant contributions to the literature on the history and scope of the CLT; relevant publications include, but are not limited to, the following:

\begin{itemize}

\item

\textbf{\textit{(Applications)}} \cite{glynn1986central} and \cite{kwak2017central};

\item

\textbf{\textit{(History)}} \cite{adams2009life}, \cite{fischer2010history}, \cite{hald-1998, hald-2007} and \cite{lecam-1986};

\item

\textbf{\textit{(Review)}} \cite{anderson2010central} and \cite{heyde2006central}; and 

\item

\textbf{\textit{(Theory and Methods)}} \cite{aistleitner2010central},
\cite{barron1986entropy}, \cite{benoist2016central}, \cite{bergstrom1944central}, \cite{bolthausen1982central, bolthausen1984estimate},
\cite{brosamler1988almost}, \cite{calegari2010combable}, \cite{chopin2004central}, \cite{cushen1971quantum}, \cite{davis1995elementary}, \cite{de1987central}, \cite{diananda1955central}, \cite{dobrushin1956central}, \cite{doukhan1994functional}, \cite{giraitis1990central}, \cite{giraitis2000stationary}, \cite{gordin1969central}, \cite{hall1984central}, \cite{hannan1979central}, \cite{hilhorst2010note}, \cite{hoeffding1948central}, \cite{hoeffding1951combinatorial}, \cite{hoffmann1976law}, \cite{johnson2004information}, \cite{jones2004markov}, \cite{kinnebrock2008note}, \cite{klartag2007central}, \cite{linnik1959information}, \cite{liverani1996central}, \cite{newman1983general}, \cite{owen1992central}, \cite{peligrad1997central}, \cite{pollard1982central}, \cite{portnoy1986central}, \cite{rosenblatt1956central}, \cite{sazonov1968multi}, \cite{schatte1988strong}, \cite{stute1995central}, \cite{trotter1959elementary}, and \cite{von1965convergence}.

\end{itemize}

\section{A short history of the CLT} \label{s:history-1}

\subsection{Jacob Bernoulli and the \textit{Weak Law of Large Numbers}} \label{s:wlln-1}

The CLT story actually begins with the \textit{Weak Law of Large Numbers} (WLLN), with the work of Jacob Bernoulli (1654--1705); see, e.g., \cite{hald-2007} for additional details. Prior to the investigations in probability undertaken by this member of the remarkable Bernoulli family, games of chance with equally-likely outcomes (based on the original Pascal-Fermat (1654) definition of probability; see, e.g., \cite{weisberg-2014}) were regarded as the only real-world examples that could be studied probabilistically. One of Bernoulli's revolutionary insights was that this narrow focus could be extended fruitfully to the study of uncertain events in what he called ``civil, moral and economic affairs,'' including applications in such fields as demography, insurance, and meteorology. As \cite{hald-2007} puts is,

\begin{quote}

Bernoulli refers to the well-known empirical fact that the relative frequency of an event, calculated from observations taken under the same circumstances, becomes more and more stable with an increasing number of observations. Noting that the statistical model for such observations is the Binomial distribution, Bernoulli asks the fundamental question: \textit{Does the relative frequency derived from the Binomial have the same property as the empirical relative frequency?} He proves that this is so and concludes that we may then extend the application of probability theory from games of chance to other fields where stable relative frequencies exist. \textit{[emphasis added]}

\end{quote}

Bernoulli's proof (\cite{bernoulli-1713}), which applies only to what we now call IID Bernoulli trials, took him more than 20 years to complete; it was published posthumously (in book form, with the WLLN proof occupying more than 200 pages) in 1713 on his behalf by his nephew Nicolaus Bernoulli. Today we summarize the WLLN as follows: \vspace*{-0.35in}

\begin{quote}

\begin{theorem} \label{t:wlln-1}

\textbf{(WLLN)} Using the same notation as in \textbf{Theorem \ref{t:clt-on-cdf-scale-1}}, and under the same assumptions, 
\begin{equation} \label{e:wlln-1}
\bar{ Y }_n \stackrel{ P }{ \rightarrow } \mu \ \ \ \textrm{as} \ \ \ n \rightarrow \infty \, ; \ \ \ \textrm{i.e., for all $\epsilon > 0$,} \ \ \ \lim_{ n \rightarrow \infty } P ( | \bar{ Y }_n - \mu | \le \epsilon ) = 1 .
\end{equation}

\end{theorem}

\end{quote}

In the notation of \textbf{Theorem \ref{t:clt-on-cdf-scale-1}}, let the $Y_i$ be IID Bernoulli$( p )$ random variables; Bernoulli succeeds in providing, for any $\epsilon > 0$, a lower bound for $P ( | \bar{ Y }_n - p | \le \epsilon )$ that approaches 1 as $n \rightarrow \infty$. His bound was rather loose: as an example of his results he shows, with $p = 0.6$, that
\begin{equation} \label{e:bernoulli-1}
P ( 0.58 \le \bar{ Y }_n \le 0.62 ) > \frac{ 1000 }{ 1001 } \ \ \ \textrm{for all} \ \ \ n \ge 25,500 \, ,
\end{equation}
whereas a CLT-based calculation gives $n \ge$ 6,498. Today we can prove the WLLN for \textit{any} $Y_i$ with finite variance in a few lines using Chebyshev's inequality, but that result (first published in \cite{chebyshev-1867}) was of course not available to Bernoulli.

\subsection{Abraham de Moivre and the Normal Approximation to the Binomial} \label{s:normal-approximation-1}

Bernoulli showed in the early 1700s that $\bar{ Y }_n$ converges in probability to $\mu$ for IID Bernoulli $Y_i$, but he did not have time in his relatively short life of 50 years to carefully answer the obvious next question: at what precise rate does this convergence occur? The first progress on this problem, also for Bernoulli trials, was provided by Abraham de Moivre (1667--1754), initially in an unpublished pamphlet (\cite{de-moivre-1733}) and then in the second edition of his remarkable book (\cite{de-moivre-1738}), \textit{The Doctrine of Chances}. It seems \citep{hald-2007} that de Moivre knew about Bernoulli's WLLN but offered two proofs of it himself anyway (also for IID Bernoulli trials) on the way to deriving two Normal approximations to the Binomial$( n, p )$ distribution, one of which we briefly examine here. 

In the notation of \textbf{Theorem \ref{t:wlln-1}}, de Moivre was interested in quantifying $P ( | \bar{ Y }_n - \mu | \le \epsilon )$ for IID Bernoulli$( p )$ trials $Y_i$. Setting $S_n \triangleq \sum_{ i = 1 }^n Y_i$ with observed value $s_n \triangleq \sum_{ i = 1 }^n y_i$, de Moivre regarded $np$ as an integer for simplicity and focused on
\begin{equation} \label{e:binomial-approximation-1}
P ( | S_n - n p | \le d ) = P \left( | \bar{ Y }_n - p | \le \frac{ d }{ n } \right) = 
\sum_{ s_n = a_n }^{ b_n } P ( S_n = s_n ) =  \sum_{ s_n = a_n }^{ b_n } \left( \begin{array}{c} n \\ s_n \end{array} \right) p^{ s_n } \, ( 1 - p )^{ n - s_n } \, ,
\end{equation}
in which $a_n = \max \{ 0, n p - d \}$, $b_n = \min \{ n p + d, n \}$, and $d = 0, 1, \dots, \max \{ n p, n ( 1 - p ) \}$ (note that Bernoulli's $\epsilon$ corresponds to de Moivre's $\frac{ d }{ n }$). de Moivre made three main contributions:

\begin{itemize}

\item[(1)]

He wanted to derive an asymptotic approximation to $\ln [ P ( S_n = s_n ) ]$, and it became evident that to do so he needed an approximation to $\ln ( n ! )$, so he derived one:
\begin{equation} \label{e:stirling-1}
\ln ( n ! ) \doteq \frac{ 1 }{ 2 } \ln ( 2 \pi n ) + n \ln ( n ) - n
\end{equation}
(this is of course now called \textit{Stirling's approximation}, although de Moivre obtained it independently of, and at the same as, \cite{stirling-1733});

\item[(2)]

He used equation (\ref{e:stirling-1}) to arrive at his desired large-sample approximation to $\ln [ P ( S_n = s_n ) ]$, namely
\begin{eqnarray} \label{e:binomial-approximation-2}
\ln [ P ( S_n = s_n ) ] & \doteq & \left( n + \frac{ 1 }{ 2 } \right) \ln ( n ) + s_n \ln ( p ) + ( n - s_n ) \ln ( 1 - p ) - \frac{ 1 }{ 2 } \ln ( 2 \pi ) \nonumber \\ & & \hspace*{0.25in} - \ln ( s_n ) - \left( s_n - \frac{ 1 }{ 2 } \right) \ln ( s_n - 1 ) \nonumber \\
& & \hspace*{0.25in} - \left( n - s_n + \frac{ 1 }{ 2 } \right) \ln ( n - s_n + 1 ) \, , 
\end{eqnarray}
in which $s_n = n p + d$ and $d = o ( n )$; and

\item[(3)]

He then found to his dismay that substituting an exponentiated version of (\ref{e:binomial-approximation-2}) into (\ref{e:binomial-approximation-1}) produced nothing tractable, so he formulated a different plan to achieve his main goal, in two steps:

\begin{itemize}

\item[(i)]

To simplify his life de Moivre looked carefully at the symmetric Binomial$( n, p )$ PMF, for which $p = \frac{ 1 }{ 2 }$; he noted that you can readily get back and forth from the symmetric case to all other Binomial distributions via the equality
\begin{eqnarray} \label{e:binomial-1}
P ( S_n = s_n ) & = & \left( \begin{array}{c} n \\ s_n \end{array} \right) p^{ s_n } \, ( 1 - p )^{ n - s_n } \nonumber \\
& = & \left[ \left( \begin{array}{c} n \\ s_n \end{array} \right) \frac{ 1 }{ 2^n } \right] \, \bigg\{ ( 2 \, p )^{ s_n } \left[ 2 \, ( 1 - p ) \right]^{ n - s_n } \bigg\} \, .
\end{eqnarray}
Again for simplicity he made the sample size even by setting $n = 2 \, m$ and examined the height of the symmetric Binomial PMF to the left and right of the tallest spike, namely $P \left( S_n = m + d \given p = \frac{ 1 }{ 2 } \right) = \left( \begin{array}{c} 2 \, m \\ m + d \end{array} \right) \frac{ 1 }{ 2^m }$ for $| d | = 0, 1, \dots, m$ and $m = 1, 2, \dots$; and

\item[(ii)]

He obtained results that have a ring of familiarity to us today, namely
\begin{equation} \label{e:binomial-2}
P \left( S_n = m + d \left| p = \frac{ 1 }{ 2 } \right. \right) \doteq \frac{ 1 }{ \sqrt{ \pi m } } \exp \left( - \frac{ d^2 }{ m } \right)
\end{equation}
and
\begin{equation} \label{e:binomial-3}
P ( S_n = n p + d ) \doteq \frac{ 1 }{ \sqrt{ 2 \pi n p ( 1 - p ) } } \exp \left[ - \frac{ d^2 }{ 2 n p ( 1 - p ) } \right] \, , \ \ \ d = O ( \sqrt{ n } ) \, ,
\end{equation}
which can be written in the notation of this paper as
\begin{equation} \label{e:binomial-4}
\sqrt{ n p ( 1 - p ) } \, P ( S_n = s_n ) \stackrel{ . }{ \sim } \phi \left( \frac{ s_n - n p }{ \sqrt{ n p ( 1 - p ) } } \right) \, .
\end{equation}
Equation (\ref{e:binomial-2}) appears to be the first time in history that the kernel 
$\exp \left( - \frac{ d^2 }{ m } \right)$ of the standard Normal PDF arose in the mathematics of probability theory. de Moivre now observed that as $n \rightarrow \infty$ the sum in equation (\ref{e:binomial-approximation-1}) will be well approximated by an integral, leading immediately to 
\begin{equation} \label{e:binomial-5}
P ( | S_n - n p | \le d ) \doteq \Phi ( z ) - \Phi ( - z ) \ \ \ \textrm{with} \ \ \ z = \frac{ d }{ \sqrt{ n p ( 1 - p ) } } \, .
\end{equation}
He used a combination of series expansion and numerical integration to obtain quantitative results based on $\Phi ( z )$, thereby achieving his goal.

\end{itemize}

We have numerically examined the approximations in equations (\ref{e:binomial-2}--\ref{e:binomial-5}) for $n$ from 10 to 100 and $p$ from 0.01 to 0.5. Because de Moivre did not account for skewness in the Binomial with $p \ne 0.5$, his PDF-scale approximations behave badly for small to moderate $n$ when $0.01 \le p \le 0.1$ (and therefore of course also for $0.9 \le p \le 0.99$), but equations (\ref{e:binomial-2}--\ref{e:binomial-4}) yield highly accurate results even for small $n$ when $0.1 \le p \le 0.9$. However, equation (\ref{e:binomial-5}) produces terrible approximations on the CDF scale even for $( n, p ) = ( 100, 0.5 )$ (see column 3 in Table \ref{t:de-moivre-1}), because de Moivre also did not include a \textit{continuity correction}. When $d$ in the expression $z = \frac{ d }{ \sqrt{ n p ( 1 - p ) } }$ is replaced by $( d + 0.5 )$, his CDF approximation (column 4 in the Table) is excellent. This \textit{continuity correction} to de Moivre's work was only introduced into the literature much later, by Augustus \cite{de-morgan-1838}.

\end{itemize}

\begin{table}[t!]

\centering

\caption{\textit{Exact Binomial CDF values for $( n, p ) = ( 100, 0.5 )$, together with de Moivre's Normal approximation without and with continuity correction (CC).}}

\bigskip

\begin{tabular}{c||c|c|c}

& Exact \\

& Binomial & \multicolumn{2}{c}{Approximation (\ref{e:binomial-5})} \\ \cline{3-4}

$d$ & CDF Value & Without CC & With CC \\

\hline

0 & 0.0796 & 0.0000 & 0.0797 \\   

1 & 0.2356 & 0.1585 & 0.2358 \\

2 & 0.3827 & 0.3108 & 0.3829 \\

3 & 0.5159 & 0.4515 & 0.5161 \\

4 & 0.6318 & 0.5763 & 0.6319 \\

5 & 0.7287 & 0.6827 & 0.7287 \\

6 & 0.8067 & 0.7699 & 0.8064 \\

7 & 0.8668 & 0.8385 & 0.8664 \\

8 & 0.9114 & 0.8904 & 0.9109 \\

9 & 0.9431 & 0.9281 & 0.9426 \\

\end{tabular}

\label{t:de-moivre-1}

\end{table}

You can see that de Moivre was only a short distance away from the CLT for IID Bernoulli trials, but interestingly he didn't make that small leap; it was left to Laplace, about 80 years later, to provide the next chapter in the story.

\subsection{Pierre Simon de Laplace and the First CLT} \label{s:clt-1}

To obtain the first valid CLT in history, as related by \cite{hald-2007}, \citet{laplace-1810} adopts a remarkable proof technique: in a setting in which the $n$ random variables $\{ Y_j, j = 1, \dots, n \}$ are IID from a common distribution, in our contemporary language Laplace \textit{creates} the idea of the \textit{characteristic function} (CF) $\psi_{ Y_j } ( t ) \triangleq E \left( e^{ i \, t \, Y_j } \right)$ (another historical first) and uses this to get the CLT in a special case. As was customary at the time, his proof would be regarded today as rather more like a proof sketch; for example, he assumes without comment that all of the moments $\mu_r^\prime = E( Y_j^r )$ of $Y_j$ (for $r = 1, 2, \dots$) exist and are finite, allowing him to expand the CF as
\begin{equation} \label{e:characteristic-function-1}
\psi_{ Y_j } ( t ) \triangleq E \left( e^{ i \, t \, Y_j } \right) = 1 + i \, \mu_1^\prime \, t - \mu_2^\prime \, \frac{ t^2 }{ 2 \, ! } + \dots \, .
\end{equation}
He now uses the three-term Taylor expansion of $\ln ( 1 + x )$ about $( x = 0 )$, namely 
$\ln ( 1 + x ) \doteq x - \frac{ x^2 }{ 2 \, ! }$, applied to both sides of (\ref{e:characteristic-function-1})
to obtain
\begin{equation} \label{e:characteristic-function-2}
\ln \psi_{ Y_j } ( t ) \doteq i \, \mu_1^\prime \, t - \sigma^2 \frac{ t^2 }{ 2 } \, ,
\end{equation}
in which (as usual) the variance is $\sigma^2 = V ( Y_j ) = \mu_2^\prime - ( \mu_1^\prime )^2$. Considering the sum $S_n = \sum_{ j = 1 }^n Y_j$ of the IID $Y_j$, Laplace now notes that $\psi_{ S_n } ( t ) = E \left( e^{ i \, t \, S_n } \right) = \left[ \psi_{ Y_j } ( t ) \right]^n$ so that on the log scale the CF of $S_n$ is approximately
\begin{equation} \label{e:characteristic-function-3}
\ln \psi_{ S_n } ( t ) \doteq n \left( i \, \mu_1^\prime \, t - \sigma^2 \frac{ t^2 }{ 2 } \right) \, ,
\end{equation}
and his next challenge is to solve an inverse problem: figuring out the distribution of $S_n$ from its CF. Assuming (to make progress) that the $Y_j$ are discrete, taking integer values $k \in \mathbb{ Z } = \{ \dots, -2, -1, 1, 2, \dots \}$ with (nonnegative) probabilities $p_k$ summing to 1, Laplace first computes 
\begin{equation} \label{e:characteristic-function-4}
\psi_{ Y_j } ( t ) = E \left( e^{ i \, t \, Y_j } \right) = \sum_{ k = - \infty }^\infty p_k \, \exp ( i t k ) \, .
\end{equation}
Recalling that
\begin{equation} \label{e:characteristic-function-5}
\frac{ 1 }{ 2 \, \pi } \int_{ - \pi }^\pi \exp ( i t k ) \, dt = \frac{ 1 }{ 2 \, \pi } \int_{ - \pi }^\pi [ \cos ( t k ) + i \sin ( t k ) ] \, dt = \left\{ \begin{array}{ccc} 1 & \textrm{if} & k = 0 \\ 0 & & k \ne 0 \end{array} \right\} \, ,
\end{equation}
Laplace gets what we would now call a Fourier inversion formula\footnote{It appears that Laplace's CLT work involving Fourier inversion formulas anticipated \cite{fourier-1822} by about 12 years.} :
\begin{equation} \label{e:characteristic-function-6}
\frac{ 1 }{ 2 \, \pi } \int_{ - \pi }^\pi \exp ( - i t k ) \, \psi_{ Y_j } ( t ) \, dt = \frac{ 1 }{ 2 \, \pi } \int_{ - \pi }^\pi \exp ( - i t k ) \left[ \sum_{ k = - \infty }^\infty p_k \, \exp ( i t k ) \right] \, dt = p_k \, .
\end{equation}
From this he can approximate the PMF of $S_n$ as follows. Since the $Y_j$ have support set $\mathbb{ Z }$, $S_n$ has the same support set, and --- in parallel with equation (\ref{e:characteristic-function-4}) --- we can write
\begin{equation} \label{e:characteristic-function-7}
\psi_{ S_n } ( t ) = E \left( e^{ i \, t \, S_n } \right) = \sum_{ \ell = - \infty }^\infty q_\ell \, \exp ( i t \ell ) \, ,
\end{equation}
in which the (nonnegative) probabilities $q_\ell$ sum to 1. In parallel with equation (\ref{e:characteristic-function-6}) this now leads to
\begin{equation} \label{e:characteristic-function-8}
\frac{ 1 }{ 2 \, \pi } \int_{ - \pi }^\pi \exp ( - i t \ell ) \, \psi_{ S_n } ( t ) \, dt = \frac{ 1 }{ 2 \, \pi } \int_{ - \pi }^\pi \exp ( - i t \ell ) \left[ \sum_{ \ell = - \infty }^\infty q_\ell \, \exp ( i t \ell ) \right] \, dt = q_\ell \, .
\end{equation}
Now Laplace chooses an $s$ of order $O ( n )$, sets $\ell = n \mu_1^\prime + s$, and exponentiates both sides of equation (\ref{e:characteristic-function-3}) to obtain
\begin{eqnarray} \label{e:characteristic-function-9}
q_\ell & = & P ( S_n = \ell = n \mu_1^\prime + s ) \doteq \frac{ 1 }{ 2 \, \pi } \int_{ - \pi }^\pi \exp \left[ - i t ( n \mu_1^\prime + s ) + n \left( i \, \mu_1^\prime \, t - \sigma^2 \frac{ t^2 }{ 2 } \right) \right] dt \nonumber \\
& = & \frac{ 1 }{ 2 \, \pi } \int_{ - \pi }^\pi \exp \left( - i t s - n \sigma^2 \frac{ t^2 }{ 2 } \right) dt \doteq \frac{ 1 }{ \sigma \sqrt{ 2 \pi n } } \exp \left( - \frac{ s^2 }{ 2 \, n \, \sigma^2 } \right) \, ,
\end{eqnarray}
the last step involving a further approximation for large $n$. Laplace concludes from this (in our current language) that $( S_n - n \mu_1^\prime )$ and $\bar{ Y }_n$ are asymptotically Normal$( 0, n \sigma^2 )$ and Normal$\left( \mu_1^\prime, \frac{ \sigma^2 }{ n } \right)$, respectively; in other words, he has offered a proof sketch (which can, in fact, be made rigorous) of the CLT for discrete random variables with finite variance. 

% hald (1750-1930) p. 314 laplace actually went from -a to +a^\prime not
% - infinity to infinity

Laplace attempted to extend his result to continuous random variables by a further approximation argument, but (as \cite{hald-2007} notes) Laplace's proof sketch could not be made rigorous; fourteen years later, Simeon de \cite{poisson-1824} patched up Laplace's flawed proof, thereby establishing the first CLT for continuous random variables with finite range of support. 

\subsection{The CLT since 1827} \label{s:clt-since-1827-1}

Here's a short list of the highlights in research on the CLT subsequent to Poisson's (1824) proof.

\begin{itemize}

\item

\cite{hald-1998} summarizes the progress on the CLT up to 1857 by noting that the work fell short of our contemporary understanding in three ways: to that point in time all authors had failed to

\begin{itemize}

\item[(1)]

extend the CLT to continuous individual observations $Y_j$ with infinite support;

\item[(2)]

identify precise moment conditions under which the \textbf{Theorem} is true; and

\item[(3)]

evaluate the remainder terms in their series expansions with sufficient accuracy to pin down the rate of convergence of the distribution of the sample mean to the Normal distribution.

\end{itemize}

\item

These deficits were removed in the period from about 1870 to 1912 by the Russian mathematicians Pafnuty \citet{chebyshev-1887}, Andre{\u\i} Andreevich \cite{markov-1912}, and Alexei \cite{lyapunov-1901}, whose assumption of a complicated condition on the moments of the $| Y_j |$ was improved upon by the Finnish mathematician Jarl Waldemar \cite{lindeberg-1920} in the following manner: \vspace*{-0.35in}

\begin{quote}

\begin{theorem} \label{t:lindeberg-1}

\textbf{Lindeberg} Let $\{ Y_i, i = 1, \dots, n \}$ be independent (but not necessarily identically distributed) random variables, with $E ( Y_i ) = \mu_i$ and $V ( Y_i ) = \sigma_i^2$; assume that all of the $\{ \mu_i, i = 1, \dots, n \}$ and $\{ \sigma_i^2, i = 1, \dots, n \}$ exist and are finite, and define $s_n^2 \triangleq \sum_{ i = 1 }^n \sigma_i^2$. Suppose that for all $\epsilon > 0$
\begin{equation} \label{e:lindeberg-condition-1}
\lim_{ n \rightarrow \infty } \frac{ 1 }{ s_n^2 } \sum_{ i = 1 }^n E \left[ ( Y_i - \mu_i )^2 I ( | Y_i - \mu_i | > \epsilon \, s_n ) \right] = 0 \, ,
\end{equation}
in which $I ( A ) = 1$ if the proposition $A$ is true and 0 otherwise. Then
\begin{equation} \label{e:lindeberg-condition-2}
\frac{ 1 }{ s_n } \sum_{ i = 1 }^n ( Y_i - \mu_i ) \stackrel{ D }{ \rightarrow } N ( 0, 1 ) \, .
\end{equation}

\end{theorem}

\textbf{Remark \remark.} \textbf{Theorem \ref{t:clt-on-cdf-scale-1}} in Section \ref{s:introduction-1} is a special case of Lindeberg's \textbf{Theorem} in which the $Y_i$ are IID with positive and finite variance $\sigma^2$.

% show that lindeberg's condition holds in the iid case

\end{quote}

\end{itemize}

% Further generalizations 

As a final historical note, the CLT was not given its name until \cite{polya-1920} chose that appellation; he intended that ``central'' should refer to the important role played by the CLT in probability theory, but --- as noted by Lucien \cite{lecam-1986} --- many French probabilists regard ``central'' as a description of 
``the behavior of the center of the distribution as opposed to its tails.''

% Finally,Cramer (1928, 1937, 1972) gave a rigorous proof of Edgeworth’s general asymptotic expansion and its extension to distribution functions. from hald before 1750

\section{Approaches to quantifying accuracy in applications of the CLT} \label{s:accuracy-quantification}

There is a broad and deep literature (see, e.g., \cite{dodge-2006}, \cite{lindsay-markatou-2002}, \cite{mahalanobis-1936}, and \cite{pardo-2006}, among many relevant publications) on ways to measure 
\begin{itemize}

\item[(I)]

the \textit{statistical distance}\footnote{This term is not precisely defined in the literature; it's intended to capture the intuitive idea of the closeness of two probability measures $\mu$ and $\nu$ defined on the same probability space. Many statistical distances are not metrics: some of them (a) fail to satisfy the triangle inequality (e.g., Bhattacharyya distance (equation (\ref{e:hellinger-2})) and/or (b) are not symmetric in their arguments (e.g., Kullback-Leibler divergence (equation (\ref{e:kl-1})).} between $F_{ Z_n } ( \cdot )$ and $\Phi ( \cdot )$ in \textbf{Theorem \ref{t:clt-on-cdf-scale-1}}, and

\item[(II)]

the \textit{rate} at which $F_{ Z_n } ( \cdot )$ approaches $\Phi ( \cdot )$ as $n$ grows.

\end{itemize}
We address (II) in Section \ref{s:edgeworth-cornish-fisher}. The key ideas in (I) naturally fall into three categories, based on whether $F_Y$ has a density or not.

\subsection{General $F_Y ( \cdot )$} \label{s:general-F}

When comparing two CDFs $F ( z )$ and $G ( z )$ that map\footnote{The set $\mathbb{ F } = \{ F \!: \mathbb{ R } \rightarrow [ 0, 1 ], F \textrm{ monotonic non-decreasing with limits 0 and 1 at } - \infty \textrm{ and } \infty, \textrm{ respectively}$\} of all CDFs on $\mathbb{ R }$, endowed with the KS or WKR metrics (defined below), is a bounded metric space (see, e.g., \cite{papadopoulos-2014}), making it a topological space in which the topology is defined by the open sets induced by the metric; $\mathbb{ F }$ is also isomorpic to the infinite-dimensional simplex $S^\infty$. These properties make it possible to do useful functional analysis on $\mathbb{ F }$, an activity that we do not pursue here.} from $\mathbb{ R }$ to $[ 0, 1 ]$, for any fixed $z$ it's natural to look at the Euclidean distance $d ( z ) \triangleq| F ( z ) - G ( z ) |$ between them, but as $z$ ranges from $- \infty$ to $+ \infty$ this generates an uncountably infinite collection $\mathcal{ C } ( z )$ of values; it's highly desirable to have a single real number $d^*$ that summarizes $\mathcal{ C } ( z )$ in such a way that $d^*$ is (a) close to 0 when $F$ and $G$ are ``almost the same'' and (b) 0 when they're identical. The most obvious candidates are obtained by computing (i) the supremum or (ii) the integral of $d ( z )$ over $\mathbb{ R }$; these lead directly to the following two well-known distance measures:
\begin{itemize}

\item

the \textit{Kolmogorov--Smirnov (KS)} metric
\begin{equation} \label{e:ks-1}
KS ( F, G ) \triangleq \sup_{ z \in \mathbb{ R } } | F ( z ) - G ( z ) | \, , \hspace*{0.3in} \textrm{and} 
\end{equation}

\item

the \textit{Wasserstein--Kantorovich--Rubinstein (WKR)} metric\footnote{This distance measure is based on a much more general formulation, the $p$th \textit{Wasserstein distance} $W_p ( \mu, \nu )$ between two probability measures $\mu$ and $\nu$ on a metric space $( M, d )$ with metric $d$; by the \textit{Kantorovich--Rubenstein Theorem} (see, e.g., \cite{gibbs-su-2002}), $WKR ( F, G )$ in equation (\ref{e:wkr-1}) is actually just Wasserstein distance with $p = 1$ and $( M, d ) = ( \mathbb{ R }, \textrm{Euclidean distance})$. We do not pursue this additional level of abstraction here.}, which when specialized to CDFs relevant to this paper is\footnote{When $F$ and $G$ are invertible, $WKR ( F, G )$ also equals $\int_0^1 | F^{ -1 } ( p ) - G^{ -1 } ( p ) | \, d p$ (\cite{gibbs-su-2002}).}
\begin{equation} \label{e:wkr-1}
WKR ( F, G ) \triangleq \int_\mathbb{ R } | F ( z ) - G ( z ) | \, d z \, .
\end{equation}

\end{itemize}

A variety of other statistical distance measures between two distributions have been studied, some of which are defined with sufficient abstraction that they apply to the case of general $F_Y ( \cdot )$ examined in this Section; this list includes the \textit{discrepancy metric}, \textit{total variation distance}, the \textit{L\'evy--Prokhorov metric}, and \textit{$\chi^2$ distance} (see, e.g., \cite{gibbs-su-2002}), none of which we examine here.

\subsection{Absolutely continuous $F_Y ( \cdot )$} \label{s:continuous-F}

So far in this discussion $F_{ Y } ( \cdot )$ has had no restrictions (e.g., corresponding to a discrete, continuous or mixed distribution) placed on it. If $F_{ Y } ( y )$ is absolutely continuous, so that it possesses a density $f_Y ( y )$ (with respect to Lebesgue measure), it's then meaningful (as in \textbf{Theorem 3} below) to define the density $f_{ Z_n } ( z )$ of the standardized mean $Z_n$ (see equation (\ref{e:edgeworth-1})) and to consider discrepancies between $f_{ Z_n } ( \cdot )$ and the standard Normal PDF $\phi ( \cdot )$.

% get all the references to theorems right

On the density scale, a number of measures of statistical distance have been defined and studied (see, e.g., \cite{hellinger-1909} and \cite{kullback-leibler-1951}); the following two are perhaps the most prominent:

\begin{itemize}

\item

the \textit{Hellinger metric} $H ( f, g )$ between densities $f$ and $g$, defined as
\begin{equation} \label{e:hellinger-1}
H ( f, g ) \triangleq \left\{ \frac{ 1 }{ 2 } \int_{ \mathbb{ R } } \left[ \sqrt{ f ( z ) } - \sqrt{ g ( z ) } \right]^2 \, d z \right\}^{ \frac{ 1 }{ 2 } } = \left\{ 1 - \int_{ \mathbb{ R } } \sqrt{ f ( z ) \, g ( z ) } \, d z \right\}^{ \frac{ 1 }{ 2 } } \, ,
\end{equation}
and its close cousin the \textit{Bhattacharyya distance} 
\begin{equation} \label{e:hellinger-2}
D_B ( f, g ) \triangleq - \ln BC ( f, g ) \, ,
\end{equation}
in which the \textit{Bhattacharyya coefficient} $BC ( f, g )$ is
\begin{equation} \label{e:hellinger-3}
BC ( f, g ) \triangleq \int_{ \mathbb{ R } } \sqrt{ f ( z ) \, g ( z ) } \, d z \, ,
\end{equation}
so that $H ( f, g ) = \sqrt{ 1 - BC ( f, g ) }$; \textrm{and}

\item

the \textit{Kullback-Leibler (KL) divergence} between densities $f$ and $g$, in asymmetric settings in which $f$ represents some form of ``truth'' and $g$ represents an approximation to $f$: the KL divergence \textit{from $g$ to $f$} is
\begin{eqnarray} \label{e:kl-1}
KL ( f \, || \, g ) & \triangleq & \int_{ \mathbb{ R } } f ( z ) \log \left[ \frac{ f ( z ) }{ g ( z ) } \right] \, d z \nonumber \\ & = & - \int_{ \mathbb{ R } } f ( z ) \log g ( z ) \, d z - \left[ - \int_{ \mathbb{ R } } f ( z ) \log f ( z ) \, d z \right] \, ,
\end{eqnarray}
in which
\begin{equation} \label{e:kl-2}
CrEn ( f, g ) \triangleq - \int_{ \mathbb{ R } } f ( z ) \log g ( z ) \, d z \ \ \ \textrm{and} \ \ \ DE ( f ) \triangleq - \int_{ \mathbb{ R } } f ( z ) \log f ( z ) \, d z 
\end{equation}
are the \textit{cross entropy of $g$ relative to $f$} and the \textit{differential entropy of $f$}, respectively; $KL ( f \, || \, g )$ is also referred to\footnote{The information-theoretic ideas underlying $KL ( f \, || \, g )$ are usually attributed to Claude \cite{shannon-1948} and Solomon Kullback and Richard Leibler (1951), but the first people to see the importance of $- \sum_{ i = 1 }^I p_i \, \ln ( p_i )$ (for discrete PMFs) as a measure of disorder (and therefore information) in statistical mechanics were Ludwig \cite{boltzmann-1868} and (especially) J.~Willard \cite{gibbs-1902}.} as the \textit{relative entropy of g with respect to f}. A symmetrized and smoothed version of (\ref{e:kl-1}), the \textit{Jensen--Shannon (JS) metric}, has also been studied: letting $fg ( z ) \triangleq \frac{ 1 }{ 2 } \left[ f ( z ) + g ( z ) \right]$ denote the finite mixture of $f$ and $g$ with equal weights, the JS metric is defined to be
\begin{equation} \label{e:kl-3}
JS ( f \, || \, g ) \triangleq \sqrt{ \frac{ 1 }{ 2 } \left[ KL ( f \, || \, fg  ) + KL ( g \, || \, fg  ) \right] } \, .
\end{equation}
We provide numerical illustrations of the four metrics \{$KS ( F, G ), \ WKR ( F, G ), \ H ( f, g ),$ $JS ( f \, || \, g )$\} in Section \ref{s:case-studies} below.

\end{itemize}

\subsection{A choice suggested by the CLT itself} \label{s:metric-choice}

Motivated by the fact that the central quantity of interest in \textbf{Theorem 1} is $\left| F_{ Z_n } ( z ) - \Phi ( z ) \right|$ (see equation (\ref{e:clt-2})), in quantifying convergence in distribution of the sequence $F_{ Z_n } ( z )$ of CDFs to $\Phi ( z )$, we focus in Sections \ref{s:edgeworth-cornish-fisher} and \ref{s:case-studies} below on the Kolmogorov--Smirnov metric: consider the functions $e^* ( n, z )$ and $e ( n, y )$ defined by
\begin{equation} \label{e:ks-2}
e^* ( n, z ) \triangleq \left| F_{ Z_n } ( z ) - \Phi ( z ) \right| = \left| P \! \left( \bar{ Y }_n \le y \right) - \Phi \left( \frac{ y - \mu }{ \sigma / \sqrt{ n } } \right) \right| \triangleq e ( n, y ) \, .
\end{equation}
Note that 
\begin{itemize}

\item[(a)] 

Equation (\ref{e:ks-2}) establishes an absolute, rather than relative, accuracy standard, and 

\item[(b)]

$KS ( F, G )$ is symmetric in its two arguments, whereas --- in the context of the CLT --- there's a natural asymmetry, in which $\Phi ( \cdot )$ may be viewed as an approximation to $F_{ Z_n } ( \cdot )$ (the same comment\footnote{In spite of this, we continue to examine $KS ( F, G )$ below because of its central role in \textbf{Theorem 1}.} applies to $KL ( f_{ Z_n } \, || \, \phi )$).

\end{itemize}
Observation (a) makes it at least worth mentioning to consider a relative discrepancy measure of the form
\begin{equation} \label{e:relative-1}
\left| \frac{ \Phi ( z ) - F_{ Z_n } ( z ) }{ F_{ Z_n } ( z ) } \right| \, ,
\end{equation}
defined for all $z$ such that $F_{ Z_n } ( z ) > 0$. In our view absolute error is (substantially) more useful than relative error in the fields of applied statistics and data science in making practical CLT calculations: bounding relative error leads to enormous required sample sizes to get the extreme tails right, a criterion that we would argue is not especially relevant to day-to-day applied work (getting a nominal central probability of (e.g.) 0.999 right to within (e.g.) 0.0005 is an example of our goals in the numerical results in this paper).

For general $F_Y ( \cdot )$, we therefore concentrate in Sections \ref{s:edgeworth-cornish-fisher} and \ref{s:case-studies} on answering questions like the following: for fixed (known) $\mu$ and $\sigma$ (see \textbf{Theorem 1}),
\begin{quote}

How does the approximation error $e^* ( n, z )$ in equation (\ref{e:ks-2}) depend on $n$ and $z$? For example, 

\begin{itemize}

\item

For fixed $z$ and specified $\epsilon > 0$, what's the smallest $n^* \triangleq n^* ( z, \epsilon )$ such that $e^* ( n, z ) \le \epsilon$? and

\item

For fixed $n$, what's $\sup_{ z \in \mathbb{ R } } e^* ( n, z ) = KS [ F_{ Z_n } ( z ), \Phi ( z ) ]$?

\end{itemize}

\end{quote}

In Section \ref{s:edgeworth-cornish-fisher} we also use the $KS$--like distance measures $\left| f_{ Z_n } ( z ) - \phi ( z ) \right|$ and $\left| F_{ Z_n }^{ -1 } ( p ) - \Phi^{ -1 } ( p ) \right|$ in examining the CLT on the PDF and quantile scales, respectively.

\section{The role of the third and fourth moments of the $Y_i$} \label{s:edgeworth-cornish-fisher}

\subsection{Initial thoughts} \label{s:initial-thoughts-1}

The CLT speaks directly to the first two moments of the distribution $F_{ Y_i } ( y ) \triangleq P ( Y_i \le y )$ of the $Y_i$; it's therefore natural to explore the role of the third (skewness) and fourth (excess kurtosis) moments in quantifying $e^*( n, z )$. Remembering that the Normal distribution has 0 values for both skewness and excess kurtosis, the following Lemma and Conjecture make good sense: \vspace*{-0.35in}

\begin{quote}

\begin{lemma} \label{l:clt-always-1}

If the $Y_i$ are Normal to begin with, the CLT applies with \textbf{no} approximation error --- $e^* ( n, z ) = 0$ --- for all integer $n \ge 1$ and all $z \in \mathbb{ R }$.

\end{lemma}

\textbf{Conjecture (Version 1).} \textit{The closer $F_{ Y_i }$ is to Normality to begin with, the smaller $n$ needs to be to enjoy a small approximation error from the CLT.}

\end{quote}

One way to measure the closeness of $F_{ Y_i }$ to $N \! \left( \mu, \sigma^2 \right)$ is to calculate the
\begin{equation} \label{e:skewness-kurtosis-1}
\textrm{skewness} ( Y_i ) \triangleq \lambda = E \left( \frac{ Y_i - \mu }{ \sigma } \right)^3 \ \ \ \textrm{and} \ \ \ \textrm{excess kurtosis} ( Y_i ) \triangleq \eta = E \left( \frac{ Y_i - \mu }{ \sigma } \right)^4 - 3 \, ,
\end{equation}
assuming that these two quantities exist and are finite, leading to
\begin{quote}

\textbf{Conjecture (Version 2).} \textit{The closer $\lambda$ and $\eta$ are to 0, the smaller $n$ needs to be for the CLT to deliver a small approximation error.}

\end{quote}

Another natural question now arises: how do the skewness and excess kurtosis of $\bar{ Y }_n$ relate to the corresponding values for $Y_i$? The answer, which was well known to statisticians --- who defined and studied \textit{cumulant generating functions} (\cite{fisher-wishart-1932}) --- in the early part of the 20th century but which appears to have been less well known lately among applied statisticians and data scientists, is as follows (see, e.g., \cite{abramowitz-stegun-1988} and \cite{mccullagh-1984}):
\begin{equation} \label{e:skewness-kurtosis-2}
\textrm{skewness} ( \bar{ Y }_n ) = \frac{ \lambda }{ \sqrt{ n } } \ \ \ \textrm{and} \ \ \ \textrm{excess kurtosis} ( \bar{ Y }_n ) = \frac{ \eta }{ n } \, .
\end{equation}
A simple answer to the question \textit{``How big does $n$ need to be?''} now suggests itself: specify targets $\Delta_S$ and $\Delta_{ EK }$ (respectively) for the skewness and excess kurtosis of $\bar{ Y }_n$ such that $| \Delta_S |$ and $| \Delta_{ EK } |$ are both close to 0, and solve for the minimum $n$ that achieves both goals. The result arising from (\ref{e:skewness-kurtosis-2}) is readily seen to be
\begin{equation} \label{e:skewness-kurtosis-3}
n^* = \Bigg \lceil \max \left\{ \left( \frac{ \lambda }{ \Delta_S } \right)^2, \left| \frac{ \eta }{ \Delta_{ EK } } \right| \right\} \Bigg \rceil \, .
\end{equation}

The problem with this solution is that $\Delta_S$ and $\Delta_{ EK }$ are not easy to specify well in practice: for example, a target of $\Delta_S = \Delta_{ EK } = \pm \, 0.01$ sounds close to 0, but this provides no direct information about how far $P \! \left( \bar{ Y }_n \le y \right)$ is from $\Phi \! \left( \frac{ y - \mu }{ \sigma / \sqrt{ n } } \right)$. Essentially the difficulty is that the skewness and excess kurtosis scales typically have no direct practical meaning, whereas the probability scale (on which $e^* ( n, z )$ resides) is immediately and directly interpretable.

A theoretically interesting result (a) relevant to the error $e^* ( n, z )$ made by the CLT approximation and (b) involving the third absolute moment of the $Y_i$ was obtained independently by the American mathematician Andrew \cite{berry-1941} and the Swedish probabilist Carl-Gustav \cite{esseen-1942}. One possible statement of their result (\cite{esseen-1956}) is as follows. \vspace*{-0.35in}

\begin{quote}

\begin{theorem} \label{t:berry-esseen-1}

\textbf{(Berry-Esseen)}. Under the assumptions of \textbf{Theorem 1}, and adding the additional assumption that $\rho \triangleq E \left[ \left| \frac{ Y_i - \mu }{ \sigma  } \right|^3 \right] < \infty$, there exists a positive constant $C$ such that
\begin{equation} \label{e:berry-esseen-1}
\left| F_{ Z_n } ( z ) - \Phi ( z ) \right| \le \frac{ C \, \rho }{\sqrt{ n } } \quad \textrm{for all } n = 1, 2, \dots \ \ \textrm{and for all } z \in \mathbb{ R } \, .
\end{equation}

\end{theorem}

\end{quote}
\textbf{Theorem \ref{t:berry-esseen-1}} shows that an upper limit on the Euclidean distance between $F_{ Z_n } ( z )$ and $\Phi ( z )$ can be found that is \textit{uniform} in $z$ and that holds for \textit{all} positive integers $n$. This goes beyond the uniform-in-$z$ convergence of $F_{ Z_n }$ to $\Phi$ mentioned in \textbf{Theorem \ref{t:clt-on-cdf-scale-1}}, by establishing the \textit{rate} at which this uniform convergence occurs.

In his original paper \cite{esseen-1942} was unable to obtain a value of $C$ smaller than 7.5, but in the mid 1950s he (\cite{esseen-1956}) identified the distributions with the smallest possible $C$ for which (\ref{e:berry-esseen-1}) holds; for arbitrary positive $h$ these are 2--point PMFs of the form
\begin{equation} \label{e:berry-esseen-2}
f ( x ) = \left\{ \begin{array}{ccl} \frac{ \sqrt{ 10 } - 2 }{ 2 } & \textrm{for} & x = \frac{ - h \, ( 4 - \sqrt{ 10 } ) }{ 2 } \\ \frac{ 4 - \sqrt{ 10 } }{ 2 } & & x = \frac{ h \, ( \sqrt{ 10 } - 2 ) }{ 2 } \end{array} \right\} ,
\end{equation}
yielding $C = \frac{ 3 + \sqrt{ 10 } }{ 6 \sqrt{ 2 \, \pi } } \doteq 0.410$. %We examine the results of this theorem numerically in Section \ref{s:case-studies} below.

\subsection{A more refined approach: \textit{Edgeworth expansions} on the CDF scale} \label{s:refined-approach}

\subsubsection{Continuous $Y_i$} \label{s:continuous-case}

We now appeal to another idea that was well known 100 years ago in the probability and statistics community but which does not seem to have been a routine part of the core training of many applied statisticians and data scientists recently: the asymptotic expansion developed by Francis Ysidro Edgeworth in 1907 (see, e.g., \cite{wallace-1958} and \cite{kolassa-2006}) for the CDF of the standardized version of $\bar{ Y }_n$. For the moment, assume that the $Y_i$ are continuous random variables; we discuss a refinement to Edgeworth's results for discrete $Y_i$ below.

% and in the Appendix.

% in the end we may not put anything of this type in an Appendix; we'll see

Defining
\begin{eqnarray} \label{e:edgeworth-2}
A_n ( z ) & \triangleq & - \frac{ \lambda }{ 6 \, \sqrt{ n } } \, \frac{ d^2 }{ dz^2 } \, \phi ( z ) = - \frac{ \lambda \, \exp{ \left( - \frac{ z^2 }{ 2 } \right) } \, ( z^2 - 1 ) }{ 6 \, \sqrt{ 2 \, \pi \, n } } \hspace*{0.5in} \textrm{and} \\
B_n ( z ) & \triangleq & \frac{ 1 }{ 24 \, n  } \, \left[ \eta \, \frac{ d^3 }{ dz^3 } \, \phi ( z ) + \frac{ \lambda^2 }{ 3 } \, \frac{ d^5 }{ dz^5 } \, \phi ( z ) \right] \nonumber \\
& = & \frac{ \exp{ \left( - \frac{ z^2 }{ 2 } \right) } \, \Big[ 3 \, \eta \, ( z^4 - 6 \, z^2 + 3 ) \, - \lambda^2 \, z \, ( z^4 - 10 \, z^2 + 15 ) \Big] }{ 72 \, n \, \sqrt{ 2 \, \pi } } \, , \nonumber
\end{eqnarray}
Edgeworth showed that --- in powers of $n^{ - \frac{ 1 }{ 2 } }$, and for fixed known $\lambda$ and $\eta$ --- useful approximations to $F_{ Z_n } ( z )$ may be obtained from the entries in Table \ref{t:edgeworth-1}.

\begin{table}[t!]

\centering

\caption{\textit{Three approximations on the CDF scale to $F_{ Z_n } ( z )$ based on the CLT, with accuracy increasing as you move from the top row down to the bottom row in the Table.}}

\bigskip

\begin{tabular}{c|c}

Approximation & Approximation \\

to $F_{ Z_n } ( z )$ & Accuracy \\ \cline{1-2}

$\Phi ( z )$ & $O ( 1 )$ \\

$\Phi ( z ) + A_n ( z )$ & $O \! \left( n^{ - \frac{ 1 }{ 2 } } \right)$ \\

$\Phi ( z ) + A_n ( z ) + B_n ( z )$ & $O \left( n^{ - 1 } \right)$

\end{tabular}

\label{t:edgeworth-1}

\end{table}

Note that this Table and the equations in (\ref{e:edgeworth-2}) offer direct and quantitative support for \textbf{Version 2} of the \textbf{Conjecture} above: correcting for nonzero $\lambda$ drives the improvement in approximation error from $O ( 1 )$ to $O \! \left( n^{ - \frac{ 1 }{ 2 } } \right)$, and $\lambda^2$ and $\eta$ both come into play in achieving the further improvement to $O \left( n^{ - 1 } \right)$.

\begin{figure}[t!]

\centering

\caption{\textit{Left panel: a plot of $g ( z ) \triangleq \left[ \exp{ \left( - \frac{ z^2 }{ 2 } \right) } \, ( z^2 - 1 ) \right]^2$ for $z \in ( - 3.5, 3.5 )$. Right panel: the CDF skewness correction $A_n ( z )$, with $\lambda \doteq + 5.07$ (as in the population-sampling example in Section \ref{s:finite-population-example}); green solid curve, $n = 50$, and red dotted curve, $n = 100$. The functions in the left and right panels are similar but not identical in shape; one is a linearly rescaled version of the square of the other.}}

\vspace*{-0.25in}

\includegraphics[ scale = 0.8 ]{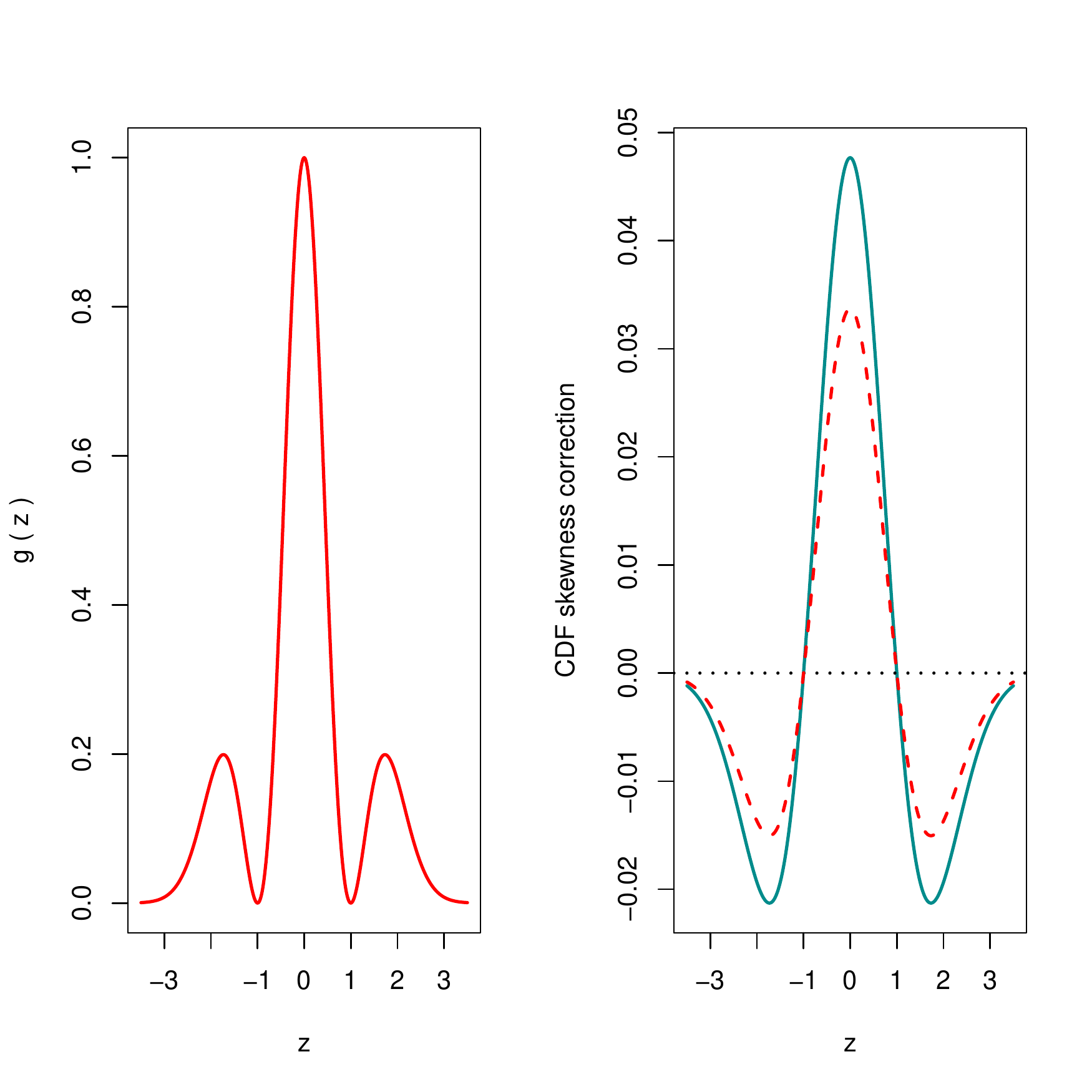}

\label{f:interesting-function-1}

\end{figure}

Defining (as above)
\begin{equation} \label{e:edgeworth-3}
e^* ( n, z ) \triangleq \left| P ( Z_n \le z ) - \Phi ( z ) \right| = \left| P \! \left( \bar{ Y }_n \le y \right) - \Phi \left( \frac{ y - \mu }{ \sigma / \sqrt{ n } } \right) \right| = e( n, y ) \, ,
\end{equation}
in which $z = \frac{ y - \mu }{ \sigma / \sqrt{ n } }$, the first equation in (\ref{e:edgeworth-2}) immediately suggests an approach to solving for $n$ to achieve an approximation error with $O \! \left( n^{ - \frac{ 1 }{ 2 } } \right)$ accuracy: set $F_{ Z_n } ( z ) \doteq \Phi ( z ) + A_n ( z )$, from which $e^* ( n, z ) \doteq \left| A_n ( z ) \right|$, and --- for a specified $\epsilon > 0$ --- solve the inequality $\left| A_n ( z ) \right| \le \epsilon$ for $n$. The result (for fixed known $\lambda$) is 
\begin{equation} \label{e:edgeworth-4}
n_3^* \triangleq n_3^* ( z, \epsilon ) \ge \Bigg \lceil \frac{ \lambda^2 \left[ \exp{ \left( - \frac{ z^2 }{ 2 } \right) } \, ( z^2 - 1 ) \right]^2 }{ 72 \, \pi \, \epsilon^2 } \Bigg \rceil
\end{equation} 
(here the 3 in $n_3^*$ refers to the third moment). The interesting function
\begin{equation} \label{e:edgeworth-5}
g ( z ) \triangleq \left[ \exp{ \left( - \frac{ z^2 }{ 2 } \right) } \, ( z^2 - 1 ) \right]^2
\end{equation}
in the numerator of (\ref{e:edgeworth-4}) is plotted in the left panel of Figure \ref{f:interesting-function-1}; it (a) achieves its global maximum at $z = 0$, (b) has two other local maxima at $z = \pm \sqrt{ 3 }$, (c) equals 0 at $z = \pm 1$, and (d) goes to 0 as $| z | \rightarrow \infty$. The right panel of the Figure displays $A_n ( z )$ for two sample sizes ($n = 50$ and $100$) in the example from Section \ref{s:finite-population-example} below --- in which the data set exhibits fairly large positive skew ($\lambda \doteq + 5.07$) --- to give an idea of the magnitude of the skewness correction on the CDF scale: with $n = 50$ the $O ( 1 )$ approximation is off by almost 0.05 at $z = 0$, and this error is still more than 0.03 at the same $z$ value with $n = 100$.

Figure \ref{f:interesting-function-1} shows that, accounting only for skewness, 
\begin{itemize}

\item

the biggest error occurs at $z = 0$, for which $g ( z ) = 1$, leading to a worst-case sample size of 
\begin{equation} \label{e:edgeworth-6}
n_{ 3, max } \triangleq n_{ 3, max } ( \lambda, \epsilon ) = \Bigg \lceil \frac{ \lambda^2 }{ 72 \, \pi \, \epsilon^2 } \Bigg \rceil \, , \ \ \ \textrm{and }
\end{equation}

\item

the $O \! \left( n^{ - \frac{ 1 }{ 2 } } \right)$ correction drops to 0 at the inflection points of the standard Normal density ($| z | = 1$) and decreases monotonically to 0 for $| z | > \sqrt{ 3 } \doteq 1.73$.

\end{itemize}
Of course, if you're only interested in a tail probability, $n_{ 3, max }$ is irrelevant and potentially (far) larger than necessary; for example, a desire to get an 0.999 central probability right to within (say) 0.001 implies an interest in $z = \pm \, \Phi^{ -1 }( 0.0005 ) \doteq \pm \, 3.29$, and $g ( \pm \, 3.29 ) \doteq 0.0019 \doteq \frac{ 1 }{ 522 }$, i.e., the required sample size at $z = \pm \, 3.29$ is about 522 times smaller than at $z = 0$.

Pursuing a similar line of inquiry to achieve an approximation error with $O \left( n^{ -1 } \right)$ accuracy, you can set $F_{ Z_n } ( z ) \doteq \Phi ( z ) + A_n ( z ) + B_n ( z )$, from which $e^* ( n, z ) \doteq \left| A_n ( z ) + B_n ( z ) \right|$, and --- for a specified $\epsilon > 0$ --- find the roots (in $n$) of the equation $\left| A_n ( z ) + B_n ( z ) \right| = \epsilon$. After some substitution and simplification, it becomes clear that the most direct way to get the solution involves setting $s \triangleq \sqrt{ n }$ and solving the equation
\begin{equation} \label{e:edgeworth-7}
\epsilon^2 = U^2 \left( \frac{ V s + W }{ s^2 } \right)^2
\end{equation}
for $s$; here
\begin{eqnarray} \label{e:edgeworth-8}
U & \triangleq & \exp \left( - \frac{ z^2 }{ 2 } \right) \, , \ \ \ V \triangleq - \frac{ \lambda \, ( z^2 - 1 ) }{ 6 \, \sqrt{ 2 \, \pi } } \, , \hspace*{0.75in} \textrm{and} \nonumber \\
W & \triangleq & \frac{ 3 \, \eta \, ( z^4 - 6 \, z + 3 ) - \lambda^2 \, z \, ( z^4 - 10 \, z^2 + 15 ) }{ 72 \, \sqrt{ 2 \, \pi } } \, .
\end{eqnarray}
This leads directly to the (depressed) quartic equation
\begin{equation} \label{e:edgeworth-9}
h ( s ) \triangleq \left( \frac{ \epsilon }{ U }\right)^2 s^4 - V^2 \, s^2 - 2 \, V \, W \, s - W^2 = 0 \, ,
\end{equation}
whose four solutions (via Ferrari's method [which dates from about 1540 (!), when Ferrari was only 18 years old (!)]) are 
\begin{equation} \label{e:edgeworth-10}
s = \frac{ - \sqrt{ U } \, \sqrt{ U \, V^2 - 4 \, W \, \epsilon } \, \pm \, U \, V }{ 2 \, \epsilon } \ \ \ \textrm{and} \ \ \ \frac{ U \, V \, \pm \, \sqrt{ U } \, \sqrt{ U \, V^2 + 4 \, W \, \epsilon }  }{ 2 \, \epsilon } \, .
\end{equation}
Some of the solutions in (\ref{e:edgeworth-10}) may be complex; ignore them (they crept in artifically through the squaring process); and then set (for fixed known $\lambda$ and $\eta$) 
\begin{equation} \label{e:edgeworth-11}
s_{ 34 } = \textrm{(largest real root in (\ref{e:edgeworth-10}) in absolute value)} \ \ \ \textrm{and} \ \ \ n_{ 34 }^* \triangleq n_{ 34 }^* ( z, \epsilon ) \ge \lceil s_{ 34 }^2 \rceil
\end{equation}

\begin{figure}[t!]

\centering

\caption{\textit{The quartic function $h ( s )$ in the population-sampling example of Section \ref{s:finite-population-example}, in which $( \lambda, \eta ) \doteq ( 5.07, 33.8 )$; the real roots of $h ( s )$ are identified with dotted vertical blue lines. Left panel: $z = 0$; right panel $z = 1$.}}

\vspace*{-0.25in}

\includegraphics[ scale = 0.8 ]{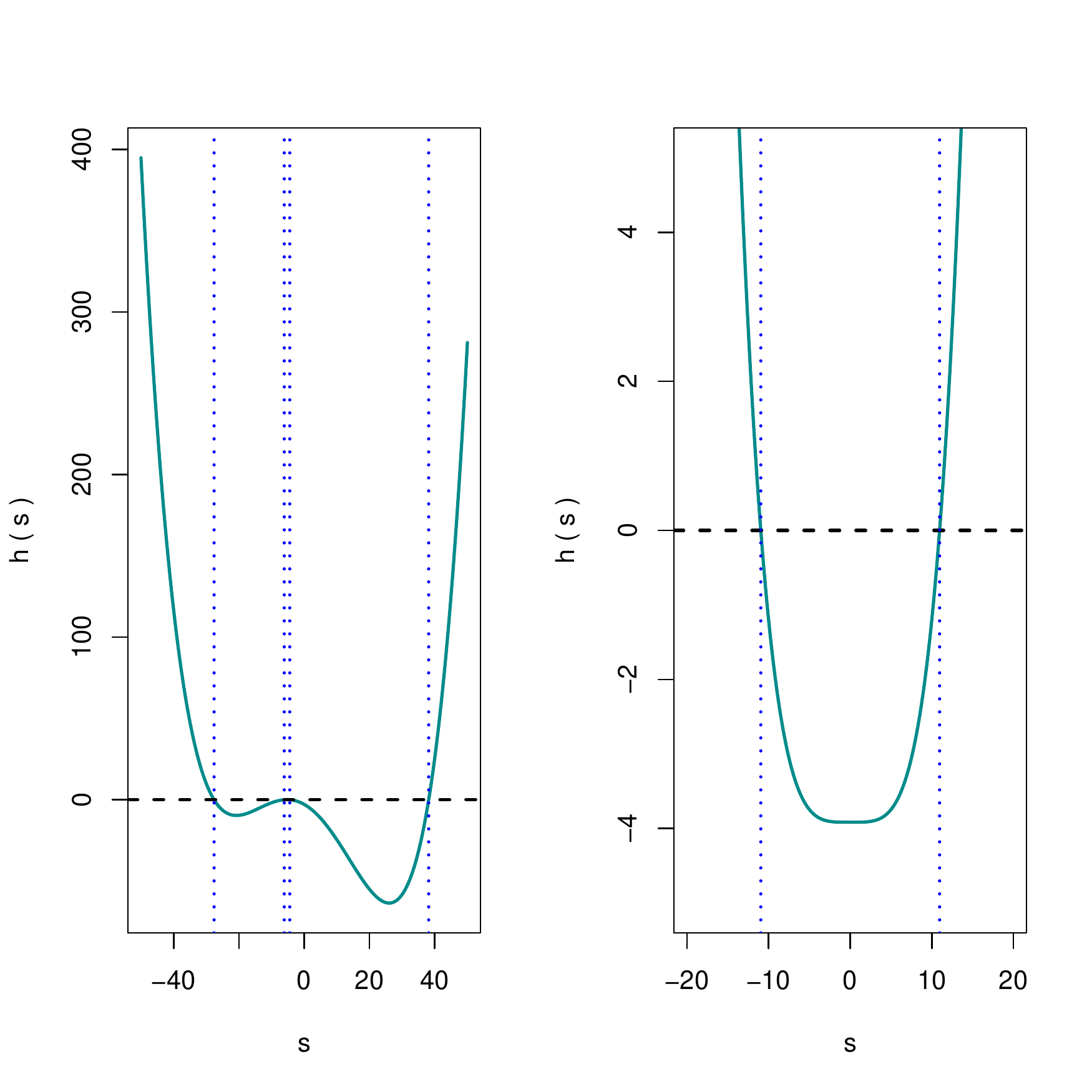}

\label{f:interesting-function-2}

\end{figure}

(here the 34 in $s_{ 34 }$ and $n_{ 34 }$ refers to the third and fourth moments). Figure \ref{f:interesting-function-2} plots the quartic function $h ( s )$ for two $z$ values in the population-sampling example of Section \ref{s:finite-population-example}; in the left panel $( z = 0 )$ there are four real roots of $h$, but only two in the right panel $( z = 1 )$. (A quartic equation may have no real roots; this could only occur here if $( U \, V^2 - 4 \, W \, \epsilon )$ and $( U \, V^2 + 4 \, W \, \epsilon )$ are both negative, which is impossible in this situation since the product $U \, V^2$ is non-negative.)

It's worth bearing in mind that all of the Edgeworth (and Cornish-Fisher [see Section \ref{s:clt-on-quantile-scale-1}]) expansions in this paper are \textit{formal} in the sense that they're not guaranteed to produce approximate \{CDF, PDF, inverse CDF\} values that respect the limits placed by probability theory on these functions (e.g., CDFs that live strictly in $[ 0, 1 ]$, PDFs that are strictly non-negative, and quantiles that are consistent with the relevant support sets). When skewness is the main source of the non-Normality of the underlying distribution, this misbehavior will only occur in one tail (e.g., the left tail with positive skew). More elaborate approximations exist that remedy this flaw in Edgeworth's method (for example, on the density scale you can derive a series approximation to $\ln \left[ f_{ Z_n } ( z ) \right]$ and exponentiate); we do not pursue this here. In Section \ref{s:finite-population-example} we examine this issue in the context of one of our case studies.

% find reference about how to keep edgeworth on cdf scale from going outside [ 0, 1 ]

\subsubsection{Discrete $Y_i$} \label{s:discrete-case}

Consider now the case that the $Y_i$ are discrete. Essentially all such random variables in practical applications\footnote{An example of a non-lattice discrete random variable would be one whose support is the set $\{ 1, \sqrt{ 2 } \}$ (you can see from the definition that the support of lattice distributions needs to be concentrated on the rational numbers). Needless to say, non-lattice distributions arise rarely, if at all, in real-world data science.} have a \textit{lattice} character: \vspace*{-0.35in}

\begin{quote}

\begin{definition} \label{d:lattice-distribution-1}

\textbf{\textit{(Lattice distribution; minimal lattice).}} \textit{A discrete random variable $Y_i$ is said to have a \textit{lattice distribution} if all of its possible values (support points) are of the form $( a + k \, h )$, where $a \in \mathbb{ R }$, $k \in \mathbb{ Z }$ and $h > 0$; $h$ is called the \textbf{span} (or \textbf{step}) of the distribution. More than one such representation always exists, involving different values of $h$; the lattice obtained from the \textit{maximum} possible $h$ --- call this span value $h_{ max }$ --- is referred to as the \textit{minimal lattice}.}

\end{definition}

\end{quote}

If the $Y_i$ are discrete, then the distribution of $Z_n$ will be discrete for all finite $n$, so that $F_{ Z_n } ( z )$ will have a step-function character; this will increase the distance between the jumpy, discontinuous CDF of $Z_n$ and the smooth continuous $\Phi ( z )$ at the jump points, although the amount of this increase will of course diminish as $n$ grows. As will be seen in what follows, it turns out that this necessitates an additional $O \! \left( n^{ - \frac{ 1 }{ 2 } } \right)$ additive correction to the approximation $[ \Phi ( z ) + A_n ( z ) ]$ to achieve the same accuracy as in the second row of Table \ref{t:edgeworth-1} when the $Y_i$ are continuous. 

To quantify this correction, here we appeal to a theorem ((3.3.2) of \cite{ibragimov-linnik-1971}), and modified to apply to our definition of $Z_n$. \vspace*{-0.35in}

\begin{quote}

\begin{theorem} \label{t:edgeworth-lattice-1}

\textbf{(Edgeworth expansion to order $O \! \left( n^{ - \frac{ 1 }{ 2 } } \right)$ for lattice random variables).} With $n$ as a positive integer, let $\{ Y_i, \, i = 1, \dots, n \}$ be IID random variables with mean $\mu$, variance $\sigma^2 > 0$, and skewness $\lambda$ (see equation (\ref{e:skewness-kurtosis-1})), all of which are assumed to exist and to be finite. Suppose that the $Y_i$ have a discrete distribution in which the support points are as in the \textit{minimal} lattice distribution identified in \textbf{Definition \ref{d:lattice-distribution-1}}, namely $\{ a + k \, h_{ max } \}$ with $h_{ max }$ maximal; set
\begin{equation} \label{e:clt-on-density-scale-2}
\bar{ Y }_n = \frac{ 1 }{ n } \sum_{ i = 1 }^n Y_i \, , \ \ Z_n \triangleq \frac{ \bar{ Y }_n - \mu }{ \sigma / \sqrt{ n } } \, , \ \ a^* \triangleq \frac{ a - \mu }{ \sigma } \ \ \ \textrm{and} \ \ h_{ max }^* \triangleq \frac{ h_{ max } }{ \sigma } \, ,
\end{equation}
and denote by $F_{ Z_n }$ the CDF of $Z_n$. Then, with $A_n ( z )$ as in equation (\ref{e:edgeworth-2}), as $n \rightarrow \infty$
\begin{equation} \label{e:lattice-correction-1}
F_{ Z_n } ( z ) = \Phi ( z ) + A_n ( z ) + \frac{ h_{ max }^* }{ \sqrt{ 2 \, \pi \, n } } \, J \left( \frac{ z \, \sqrt{ n } - n \, a^* }{ h_{ max }^* } \right) \, \exp \left( - \frac{ z^2 }{ 2 } \right) + o \! \left( n^{ - \frac{ 1 }{ 2 } } \right) \, ;
\end{equation}
here $J$ is the \textit{zig-zag (jump)} function
\begin{equation} \label{e:J-function-1}
J ( z ) = \frac{ 1 }{ \pi } \, \lim_{ \ell \rightarrow \infty} \sum_{ j = 1 }^\ell \frac{ \sin ( 2 \pi j z ) }{ j } \, .
\end{equation}

\end{theorem}

\end{quote}

\begin{figure}[t!]

\centering

\caption{\textit{$J ( z )$ for $- 3.5 \le z \le + 3.5$.}}

\vspace*{-0.15in}

\includegraphics[ scale = 0.8 ]{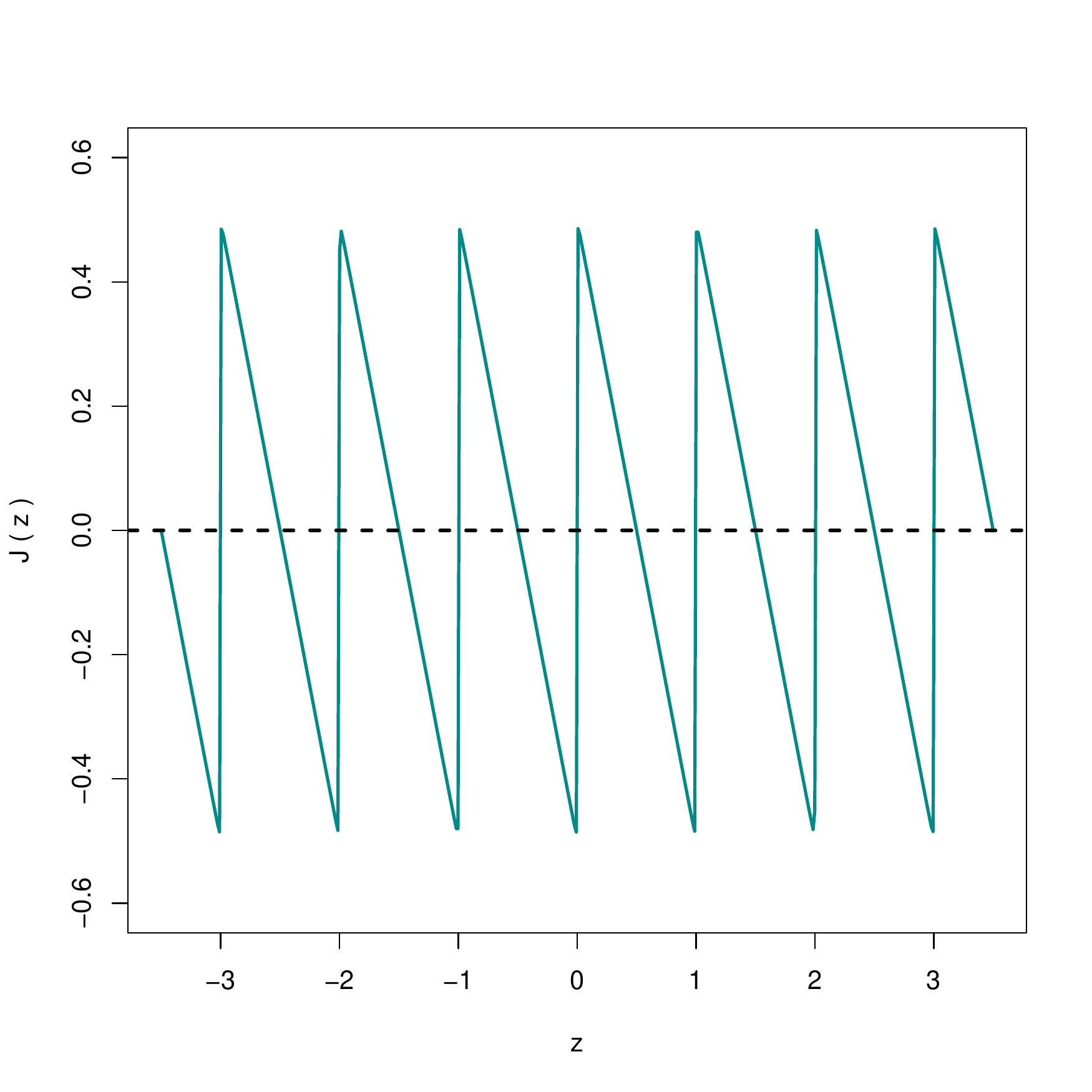}

\label{f:J-function-et-al-1}

\end{figure}

In \textbf{Theorem \ref{t:edgeworth-lattice-1}}, note that
\begin{itemize}

\item

$a$ and $h_{ max }$ are defined on the original scale of the $Y_i$ and need to be standardized to $a^*$ and $h_{ max }^*$ (since $\bar{ Y }_n$ is standardized in creating $Z_n$) before the computation in (\ref{e:lattice-correction-1}) is made;

\item

$J ( \cdot )$ is piecewise affine and periodic with period 1, taking values from $- \frac{ 1 }{ 2 }$ to $\frac{ 1 }{ 2 }$, with jump discontinuities at the integers, and with $J ( z ) = 0$ when $z$ is any integer multiple of $\pm \, \frac{ 1 }{ 2 }$; Figure \ref{f:J-function-et-al-1} is a plot of $J ( z )$ for $- 3.5 \le z \le + 3.5$. A comparison of the Fourier-series form of $J ( \cdot )$ in equation (\ref{e:J-function-1}) with its graph in Figure \ref{f:J-function-et-al-1} is surprising, and in fact there is another \textit{theoretically} equivalent representation of $J ( \cdot )$, as follows:
\begin{equation} \label{e:J-function-2}
J ( z ) = \left\{ \begin{array}{ccc} [ z ] - z + \frac{ 1 }{ 2 } & \textrm{for} & \textrm{non-integer } z > 0 \\ 0 & & \textrm{integer } z \\ \mbox{} [ z ] - z - \frac{ 1 }{ 2 } & & \textrm{non-integer } z < 0 \end{array} \right\} \, ,
\end{equation}
in which $[ z ]$ is the integer part\footnote{Until you dig deeper, it's entirely mysterious that the two different representations of $J ( z )$ in (\ref{e:J-function-1}) and (\ref{e:J-function-2}) are the same, but the mystery disappears when you think about the CLT on the characteristic function (CF) scale, which we do not pursue here; see \cite{kolassa-2006} for a lucid presentation of many of the issues addressed by this paper from the CF point of view.} of $z$. We have discovered to our dismay that, when equation (\ref{e:J-function-2}) is implemented in a software environment with finite-precision arithmetic (such as \texttt{R} or \texttt{Python}), accumulation of tiny roundoff errors can produce sharply incorrect results\footnote{We conjecture, but have not verified, that in infinite-precision (symbolic) computational environments --- such as \texttt{Mathematica}, \texttt{WolframAlpha}, or \texttt{Maple} --- the form of $J ( \cdot )$ in equation (\ref{e:J-function-2}) may be used successfully.} In practice when computing $J ( \cdot )$ with the Fourier series it's (of course) not necessary to evaluate an infinite number of terms in the sum in (\ref{e:J-function-1}) (we have had fast and accurate results with $\ell = $1,000); and

%; the Appendix contains an illustration of the failure of equation (\ref{e:J-function-2}) when implemented in \texttt{R}. 

% in the end we may not have such an appendix; think about this later

\item

The approximation accuracy provided by (\ref{e:lattice-correction-1}) can be improved to $O \! \left( n^{ - 1 } \right)$ by adding two terms, one of which is the skewness-kurtosis correction $B_n ( z )$ in equation (\ref{e:edgeworth-2}) and the other of which is an additional lattice correction term; we do not pursue this here (see \cite{kolassa-2006} or \cite{ibragimov-linnik-1971} for details).

\end{itemize}

Note further the following interesting, and crucial, point about the role played by $a$ and $a^*$ in the $J ( \cdot )$ function in \textbf{Theorem \ref{t:edgeworth-lattice-1}}: having identified the maximal standardized span as $h_{ max }^*$, it turns out that you can take \textit{any} standardized $a^*$ of the form $( a^* + k \, h_{ max }^* )$ for $k \in \mathbb{ Z }$ \textit{and the value of equation (\ref{e:lattice-correction-1}) remains the same}. The reason is that $a^*$ enters (\ref{e:lattice-correction-1}) only through the ratio $\frac{ a^* }{ h_{ max }^* }$, and the difference of any two such ratios from two different choices of $a^*$ is always an integer; this leads to the same value of the $J ( \cdot )$ function in (\ref{e:lattice-correction-1}) for all choices of $a^*$ (holding the span fixed at $h_{ max }^*$), because $J ( z + k ) = J ( z )$ for \textit{any} $z$ and \textit{any} $k \in \mathbb{ Z }$.

We illustrate the application of \textbf{Theorem \ref{t:edgeworth-lattice-1}} in Section \ref{s:roulette} below.

%, and in the Appendix we give some details on things that can go wrong in applying the theory of this section.

% in the end we may not have such an appendix; think about this later

\subsubsection{Mixed Discrete-Continuous $Y_i$} \label{s:mixed-discrete-continuous-case}

Consider now a setting in which an outcome variable of interest $Y$ is most appropriately modeled with a mixture distribution having one or more discrete components and one or more constinuous components. On the CDF scale the simplest version of this situation may be expressed as follows:
\begin{equation} \label{e:mixed-distribution-1}
F_Y ( y ) = \pi_1 \, F_Y^D ( y ) + \pi_2 \, F_Y^C ( y ) \, .
\end{equation}
Here (see equation (\ref{e:lebesgue-decomposition-1}) in Section \ref{s:introduction-1}) $0 < \pi_1 , \pi_2 < 1$ with $( \pi_1 + \pi_2 ) = 1$ and $F_Y^D$ and $F_Y^C$ are the CDFs of the discrete and continuous components, respectively. Because of the linear nature of the operation combining the two components in (\ref{e:mixed-distribution-1}), the methods of the two previous subsections ($Y_i$ discrete, $Y_i$ continuous) may be applied without the need for any new ideas. We do not delve into details in this setting in this paper.

\subsection{The CLT on the PDF scale} \label{s:clt-on-pdf-scale-1}

Central Limit Theorems are also available on the scale of PDFs, under the name \textit{local limit theorems}. Here's an example, from \cite{petrov-1975}. \vspace*{-0.35in}

\begin{quote}

\begin{theorem} \label{t:clt-on-pdf-scale}

\textbf{(CLT on the PDF scale).} With $n$ as a positive integer, let $\{ Y_i, \, i = 1, \dots, n \}$ be IID random variables with mean $\mu$ and variance $\sigma^2 > 0$, both of which are assumed to exist and to be finite, and assume that the CDF $F_{ Y_i } ( y )$ of the $Y_i$ is absolutely continuous, so that the PDF $f_{ Y_i } ( y ) \triangleq 
\frac{ d }{ d y } F_Y ( y )$ exists\footnote{Strictly speaking, as is well known, densities are unique only up to sets of Lebesgue measure 0; in this paper we're uninterested in what happens on such null sets.} for all $y \in \mathbb{ R }$; set
\begin{equation} \label{e:clt-on-density-scale-1}
\bar{ Y }_n = \frac{ 1 }{ n } \sum_{ i = 1 }^n Y_i \ \ \ \textrm{and} \ \ \ Z_n \triangleq \frac{ \bar{ Y }_n - \mu }{ \sigma / \sqrt{ n } } \, ,
\end{equation}
and denote by $f_{ Z_n } ( z )$ the density of $Z_n$. Then a necessary and sufficient condition for 
\begin{equation} \label{e:clt-on-density-scale-3}
\sup_{ z \in \mathbb{ R } } \left| f_{ Z_n } ( z ) - \phi ( z ) \right| \rightarrow 0 \ \ \ \textrm{as} \ \ \ n \rightarrow \infty
\end{equation}
is the existence of a positive $N$ such that $f_{ Z_N } ( z )$ is bounded for all $z \in \mathbb{ R }$.

\end{theorem}

\end{quote}

The necessary and sufficient condition in \textbf{Theorem \ref{t:clt-on-pdf-scale}} (a) is easy to check and (b) is essentially always true in practical applications\footnote{It's actually quite hard to create a distribution on $\mathbb{ R }$ with finite variance having the property that unboundedness of the PDF for $Y_1$ is preserved under convolution (i.e., the summation process leading to $\bar{ Y }_n$); for instance, the Beta$\left( \frac{ 1 }{ 2 }, \frac{ 1 }{ 2 } \right)$ PDF is unbounded, but already for any $n \ge 2$ the PDF of the sum of $n$ IID Beta$\left( \frac{ 1 }{ 2 }, \frac{ 1 }{ 2 } \right)$ random variables is bounded.}; in particular it's satisfied in both of our case studies in Section \ref{s:case-studies}.

Edgeworth also provided an asymptotic expansion for the probability behavior of $\bar{ Y }_n$ on the scale of its PDF, if the $Y_i$ are continuous; this second PDF expansion coheres with the CDF expansion above in the sense that differentiating the CDF expansion term-by-term yields the PDF result. Table \ref{t:edgeworth-2}
provides the analogue of Table \ref{t:edgeworth-1} on the PDF scale; here
\begin{eqnarray} \label{e:edgeworth-12}
C_n ( z ) & = & - \frac{ \lambda }{ 6 \, \sqrt{ n } } \, \frac{ d^3 }{ dz^3 } \, \phi ( z ) = \frac{ \lambda \, \exp{ \left( - \frac{ z^2 }{ 2 } \right) } \, z \, ( z^2 - 3 ) }{ 6 \, \sqrt{ 2 \, \pi \, n } } \hspace*{0.5in} \textrm{and} \\
D_n ( z ) & = & \frac{ 1 }{ 24 \, n  } \, \left[ \eta \, \frac{ d^4 }{ dz^4 } \, \phi ( z ) + \frac{ \lambda^2 }{ 3 } \, \frac{ d^6 }{ dz^6 } \, \phi ( z ) \right] \nonumber \\
& = & \frac{ \exp{ \left( - \frac{ z^2 }{ 2 } \right) } \, \Big[ 3 \, \eta \, ( z^4 - 6 \, z^2 + 3 ) + \lambda^2 \, ( z^6 - 15 \, z^4 + 45 \, z^2 - 15 ) \Big] }{ 72 \, n \, \sqrt{ 2 \, \pi } } \, . \nonumber
\end{eqnarray}
Sample size equations could now be derived for the $O \! \left( n^{ - \frac{ 1 }{ 2 } } \right)$ and $O \left( n^{ - 1 } \right)$ approximations on the PDF scale, in a manner analogous to the development in Section \ref{s:continuous-case}; we do not pursue this here, because needing to get the PDF right to within a specified accuracy target arises much less frequently in day-to-day problem-solving than targeting the CDF to produce accurate tail-area approximations.

\begin{table}[t!]

\centering

\caption{\textit{Three approximations on the PDF scale to $f_{ Z_n } ( z )$ based on the CLT, with accuracy increasing from the top row to the bottom row in the Table.}}

\bigskip

\begin{tabular}{c|c}

Approximation & Approximation \\

to $f_{ Z_n } ( z )$ & Accuracy \\ \cline{1-2}

$\phi ( z )$ & $O ( 1 )$ \\

$\phi ( z ) + C_n ( z )$ & $O \! \left( n^{ - \frac{ 1 }{ 2 } } \right)$ \\

$\phi ( z ) + C_n ( z ) + D_n ( z )$ & $O \left( n^{ - 1 } \right)$

\end{tabular}

\label{t:edgeworth-2}

\end{table}

\subsection{The CLT on the quantile (inverse CDF) scale} \label{s:clt-on-quantile-scale-1}

CLTs also exist on the quantile (inverse CDF) scale; in fact, a result of this type follows directly from \textbf{Theorem \ref{t:clt-on-cdf-scale-1}} (on the CDF scale), which may be restated as follows.

\begin{quote}

\textbf{Theorem 1$^\prime$. \textit{(reformulation of the CLT on the CDF scale)}} \textit{With $n$ as a positive integer, let $\{ Y_i, \, i = 1, \dots, n \}$ be IID random variables with mean $\mu$ and variance $\sigma^2 > 0$, both of which are assumed to exist and to be finite; set $\bar{ Y }_n = \frac{ 1 }{ n } \sum_{ i = 1 }^n Y_i$, let $Z_n$ be the standardized (mean 0, SD 1) version of $\bar{ Y }_n$ (as in equation (\ref{e:edgeworth-1}) above), and denote by $F_{ Z_n } ( z )$ the CDF of $Z_n$. Then $\sup_{ z \in \mathbb{ R } } \left| F_{ Z_n } ( z ) - \Phi ( z ) \right| \rightarrow 0$ as $n \rightarrow \infty$.}

\end{quote}

If, in addition to the assumptions in \textbf{Theorem \ref{t:clt-on-cdf-scale-1} (1$^\prime$)} the condition that the $Y_i$ are continuous is added, a basic result in functional analysis\footnote{Let $S_X$ and $S_Y$ be metric spaces and consider a sequence of functions $f_n \! \! : S_X \rightarrow S_Y$ and a uniformly continuous function $f \! : S_X \rightarrow S_Y$; suppose that the $f_n$ and $f$ have inverse functions $f_n^{ -1 }$ and $f^{ -1 }$, respectively. If the $f_n$ converge uniformly to $f$, then the $f_n^{ -1 }$ converge uniformly to $f^{ -1 }$ (see, e.g., \cite{barvinek-et-al-1991}).} then leads immediately to \vspace*{-0.35in}

\begin{quote}

\begin{theorem} \label{t:clt-on-quantile-scale-1}

\textbf{(CLT on the inverse CDF (quantile) scale).} To the assumptions in \textbf{Theorem \ref{t:clt-on-cdf-scale-1} (1$^\prime$)} add the further condition that the CDF $F_{ Y_i } = P ( Y_i \le y )$ of the $Y_i$ is absolutely continuous, from which (a) the CDF $F_{ Z_n } ( z ) = P ( Z_n \le z )$ is continuous for all real $z$ and (b) the inverse CDF $F_{ Z_n }^{ -1 } ( p )$, defined by the relation $F_{ Z_n }^{ -1 } ( p ) = z_p \textrm{ iff } F_{ Z_n } ( z_p ) = p$, is continuous for all $0 < p < 1$. Then
\begin{equation} \label{e:quantile-3}
\sup_{ p \in ( 0, 1 ) } \left| F_{ Z_n }^{ -1 } ( p ) - \Phi^{ -1 } ( p ) \right| \rightarrow 0 \ \ \ \textrm{as} \ \ \  n \rightarrow \infty \, .
\end{equation}

\end{theorem}

\end{quote}

\begin{table}[t!]

\centering

\caption{\textit{Three approximations on the inverse CDF (quantile) scale to $F_{ Z_n }^{ -1 } ( p )$ based on the CLT, with accuracy increasing as you move from the top row down to the bottom row in the Table.}}

\bigskip

\begin{tabular}{c|c}

Approximation & Approximation \\

to $F_{ Z_n }^{ -1 } ( p )$ & Accuracy \\ \cline{1-2}

$\Phi^{ - 1 } ( p )$ & $O ( 1 )$ \\

$\Phi^{ -1 } ( p ) + U_n ( p )$ & $O \! \left( n^{ - \frac{ 1 }{ 2 } } \right)$ \\

$\Phi^{ -1 } ( p ) + U_n ( p ) + V_n ( p )$ & $O \left( n^{ - 1 } \right)$

\end{tabular}

\label{t:cornish-fisher-1}

\end{table}

Edgeworth did not develop an expansion on this scale in parallel with his results on the CDF and PDF scales; it was left to the distinguished Australian statistician E.A.~Cornish and the great English statistician and geneticist R.A.~Fisher (working together), 30 years after Edgeworth's results, to develop what has since been known as the \textit{Cornish-Fisher expansion} (\cite{cornish-fisher-1938}). Table \ref{t:cornish-fisher-1} gives the analogue of Tables 1 and 2 for the inverse CDF scale; the $U_n ( p )$ and $V_n ( p )$ corrections are as in equations (\ref{e:cornish-fisher-1}), in which the $He_j ( z ) \triangleq ( - 1 )^j \exp \left( \frac{ z^2 }{ 2 } \right) \frac{ d^j }{ dz^j } \exp \left( - \frac{ z^2 }{ 2 } \right)$ are the (probabilists')\footnote{For some reason, physicists work instead with their own rescaled version $H_j ( z )$ of Hermite polynomials, in which $H_j ( z ) = 2^{ \frac{ j }{ 2 } } \, He_j ( \sqrt{ 2 } \, z )$.} Hermite polynomials:
\begin{eqnarray} \label{e:cornish-fisher-1}
U_n ( p ) & = & \frac{ \lambda }{ 6 \, \sqrt{ n } } \,  \, He_2 \! \left[ \Phi^{ -1 } ( p ) \right] = \frac{ \lambda \, \left[ \Phi^{ -1 } ( p )^2 - 1 \right] }{ 6 \, \sqrt{ n } } \hspace*{0.5in} \textrm{and} \\
V_n ( p ) & = & \frac{ \eta }{ 24 \, n } \, He_3 \! \left[ \Phi^{ -1 } ( p ) \right] - \frac{ \lambda^2 }{ 36 \, n } \, \left\{ 2 \, He_3 \! \left[ \Phi^{ -1 } ( p ) \right] + He_1 \! \left[ \Phi^{ -1 } ( p ) \right] \right\} \nonumber \\
& = & \frac{ \Phi^{ -1 } ( p ) \left\{ 3 \, \eta \, \left[ \Phi^{ -1 } ( p )^2 - 3 \right] + 2 \, \lambda^2 \left[ 5 - 2 \, \Phi^{ -1 } ( p )^2 \right] \right\} }{ 72 \, n } \, , \nonumber
\end{eqnarray}

\section{Case studies} \label{s:case-studies}

\subsection{IID sampling from a finite population} \label{s:finite-population-example}

\begin{figure}[t!]

\centering

\caption{\textit{Left panel: U.S.~family incomes in 2009, truncated at \$1 million; the data values are in units of \$1,000. Right panel: standardized income, with the standard Normal distribution superimposed (solid green curve); the blue vertical dotted line is at $z_p = + 2.576$ for $p = 0.995$, leaving a right-tail area of $( 1 - p ) = 0.005$.}}

\vspace*{-0.25in}

\includegraphics[ scale = 0.8 ]{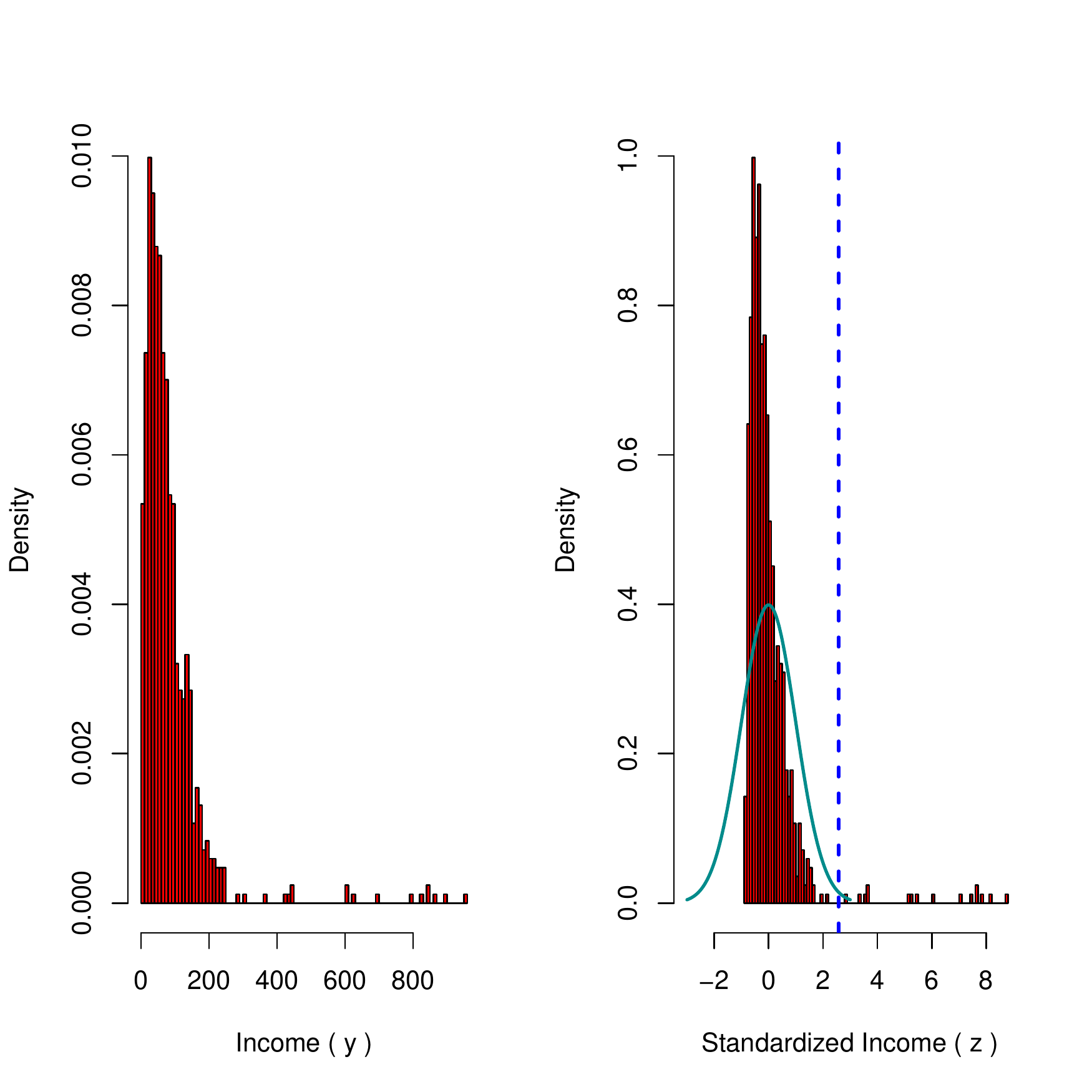}

\label{f:income-1}

\end{figure}

\begin{figure}[t!]

\centering

\caption{\textit{Histogram of $M = 10^{ 6 }$ simulated standardized $\bar{ Y }_n$ values drawn in an IID manner from the U.S.~family income population with $n = 50$; the standardized value $z_p = +2.576$ is indicated by the blue vertical dotted line. The red dotted curve, green solid curve, and black dotted curve are the $O ( 1 )$, $O \left( n^{ - \frac{ 1 }{ 2 } } \right)$, and $O \left( n^{ - 1 } \right)$ approximations, respectively, on the PDF scale.}}

\vspace*{-0.25in}

\includegraphics[ scale = 0.8 ]{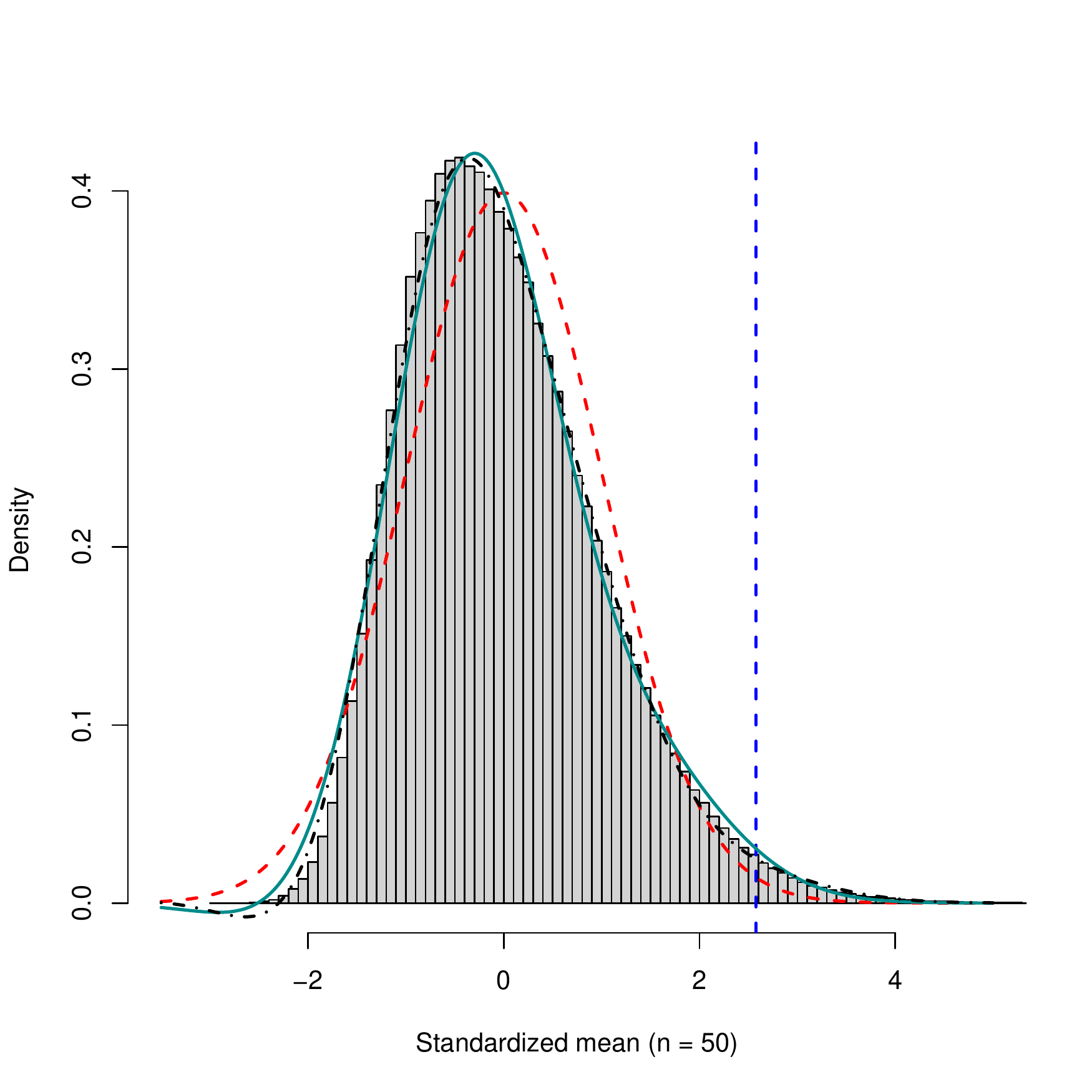}

\label{f:income-2}

\end{figure}

\subsubsection{Context} \label{s:context-1}

The left panel of Figure \ref{f:income-1} presents a histogram of an IID sample of size $N = 842$ of family incomes (in units of \$1,000) in the year 2009 from the set of all U.S.~families with income of \$1 million or less (obtained from \texttt{census.gov}); in this example we view this data set as itself a population to be sampled from in an IID manner, and we regard random draws from it as most appropriately modeled with continuous $Y_i$ (since the gaps between attainable data values --- \$1 in size --- are tiny in relation to quantities of interest, such as the typical family income [e.g., the median of this data set is about \$60k]). The population mean and SD are $( \mu, \sigma ) \doteq ( 82.88, 100.1 )$, and it's evident that substantial skewness and tail-weight are present: $( \lambda, \eta ) \doteq ( 5.070, 33.81 )$, with some observations more than 8 SDs above the mean. 

The right panel of Figure \ref{f:income-1} displays a histogram of the standardized income variable, with the standard Normal density superimposed and with the potentially interesting quantile\footnote{In view of the recent realization that a number of scientific fields, perhaps especially in the social sciences, have experienced a \textit{false-discovery crisis} (in which major studies affecting theory and methodology have failed to replicate) arising in part from the highly influential recommendation by \cite{fisher-1925} in which statistical significance may be announced with a $p$--value cutoff of 0.05, we focus here on 0.99 and 0.999 standard Normal central probability regions (we include 0.95 only for what might be termed backward compatability with the old replicability standard); thus we present numerical results for $( p, z_p ) = ( 0.975, +1.960 ), ( 0.995, +2.576 )$, and $( 0.9995, +3.291 )$. Ironically, as Table \ref{t:income-results-1} illustrates, \textit{larger} sample sizes are needed for accurate approximations with Fisher's cutpoint, using our absolute --- rather than relative --- accuracy criterion on the probability scale.} $z_p = +2.576$ (corresponding to $p = 0.995$) identified by the blue dotted line. Clearly the CLT delivers a terrible $O ( 1 )$ approximation on the density scale with $n = 1$, and it also performs badly on the CDF scale: the $O ( 1 )$ CLT approximation thinks that 0.5\% of the standardized income values should be beyond $z_{ 0.995 }$, and the actual proportion of such values is almost four times larger (about 1.90\%).

\subsubsection{The CLT in this example with $\bm{ n = 50 }$} \label{s:clt-with-n=50}

Figure \ref{f:income-2} quantifies the central limiting effect with this population data set for $n = 50$: the figure presents a histogram of $M = 10^6$ simulated $Z_{ 50 }$ values, with $z_{ 0.995 }$ again highlighted as a blue dotted vertical line. The red dotted curve is the $O ( 1 )$ CLT density result, which is clearly still unsatisfactory in all parts of the distribution with this value of $n$; the green solid and black dotted curves give the $O \left( n^{ - \frac{ 1 }{ 2 } } \right)$ and $O \left( n^{ - 1 } \right)$ approximations, respectively, using the $C_n ( z )$ and $D_n ( z )$ corrections in (\ref{e:edgeworth-12}). The Edgeworth corrections yield noticeably better results, although (a) the $O \left( n^{ - 1 } \right)$ correction hardly makes a difference on the density scale with this value of $n$ and (b) both Edgeworth improvements embarrass themselves by going slightly negative in the left tail.

\subsubsection{Avoiding negativity on the PDF scale} \label{s:avoiding-negativity-1}

\begin{figure}[t!]

\centering

\caption{\textit{Minimum $n$ to keep the $O \! \left( n^{ - \frac{ 1 }{ 2 } } \right)$ approximation on the PDF scale from going negative in the left tail when the skewness $\lambda > 0$; the cross-hairs identify the point $( \lambda, z_* ) = ( 5.07, -3.0 )$ of interest in the income case study.}}

\vspace*{-0.25in}

\includegraphics[ scale = 0.8 ]{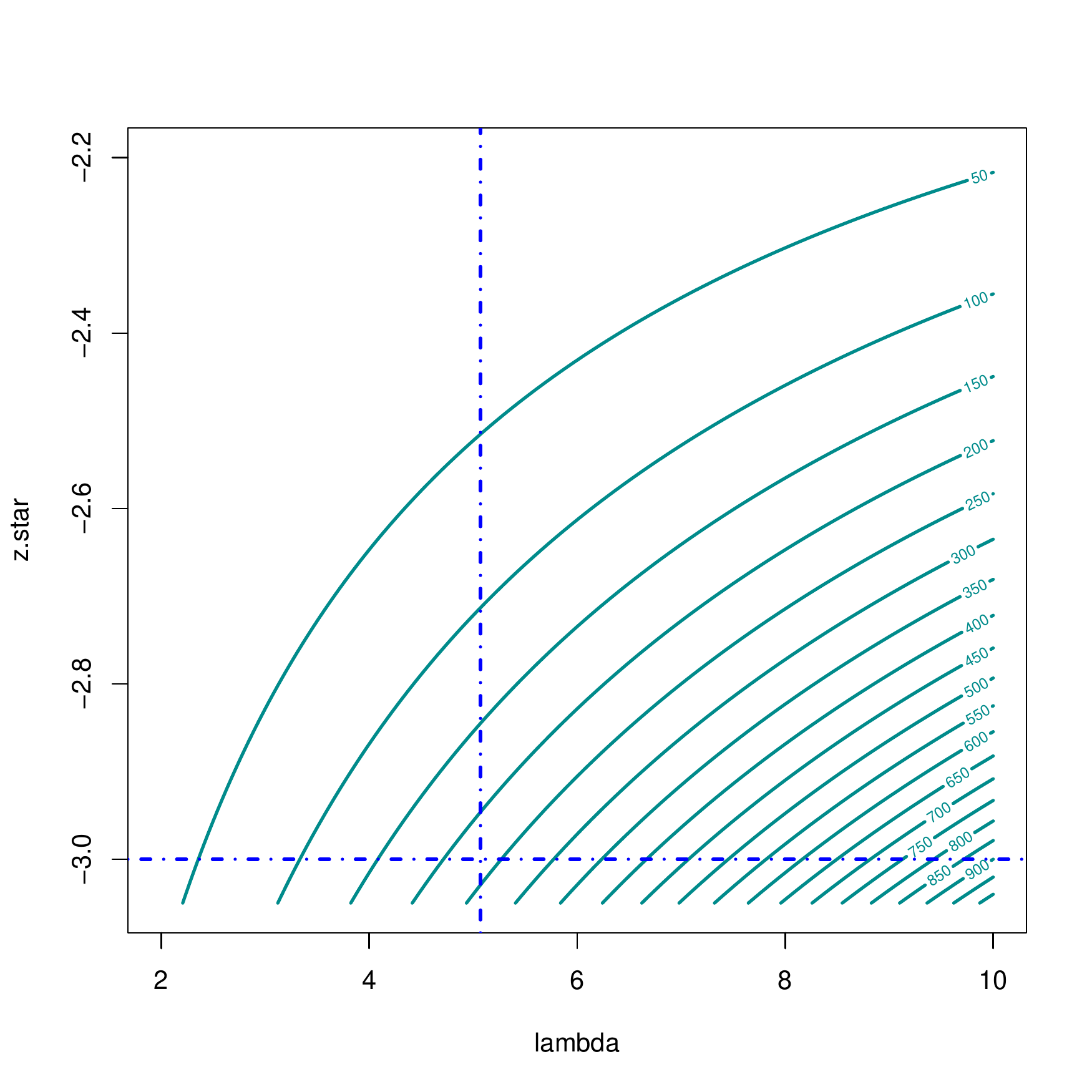}

\label{f:avoiding-negativity-1}

\end{figure}

One way to investigate the left-tail misbehavior in this case study is to examine the $O \! \left( n^{ - \frac{ 1 }{ 2 } } \right)$ approximation
\begin{equation} \label{e:avoiding-negativity-1}
f_{ Z_n } ( z ) \doteq \phi ( z ) + C_n ( z ) = \phi ( z ) \left[ 1 + \frac{ \lambda  \, z \, ( z^2 - 3 ) }{ 6 \, \sqrt{ n } } \right] \triangleq \phi ( z ) \, \tau ( n, z, \lambda )
\end{equation}
from Section \ref{s:clt-on-pdf-scale-1} and to ask under what conditions $\tau ( n, z, \lambda )$ can be prevented from going negative in the $z$ region of greatest interest. Recalling that $\lambda$ in this example is strongly positive, so that only the left tail needs attention, it's straightforward to show that the minimum $n$ such that $\tau ( n, z, \lambda ) > 0$ for all $z > z_*$ is
\begin{equation} \label{e:avoiding-negativity-2}
n^\dagger = \textrm{ceiling of } \left[ \frac{ \lambda \, z_* ( 3 - z_*^2 ) }{ 6 } \right]^2 \, .
\end{equation}
Figure \ref{f:avoiding-negativity-1} presents a contour plot of the expression for $n^\dagger$ in equation (\ref{e:avoiding-negativity-2}), across the ranges $\{ \lambda = ( 2, 10 ), z_* = ( -3.05, -2.2 ) \}$ of interest in this case study; in particular, to avoid left-tail embarrassment with $( \lambda, z_* ) = ( 5.07, -3.0 )$, it turns out that $n = 232$ or more observations are needed.

\subsubsection{Numerical results} \label{s:numerical-results-1}

With $n = 50$ the skewness and excess kurtosis of $\bar{ Y }_n$ have been driven down from $( \lambda, \eta )_{ n = 1 } \doteq ( 5.070, 33.81 )$ in the population to $( \lambda, \eta )_{ n = 50 } \doteq ( 0.7156, 0.3430 )$, in accordance with the equations in (\ref{e:skewness-kurtosis-2}) above; these measures of non-Normality are not yet (in the sense of increasing $n$) close enough to 0 for the $O ( 1 )$ CLT result to be highly accurate. For instance, row 1 of Table \ref{t:edgeworth-1} predicts that 0.5\% of the standardized means with $n = 50$ will exceed the $z_{ 0.995 } = +2.576$ value, but the actual percentage is about 1.52\%; when Table \ref{t:edgeworth-1} is applied appropriately to yield results for right-tail areas, the $O \left( n^{ - \frac{ 1 }{ 2 } } \right)$ and  $O \left( n^{ - 1 } \right)$ corrections yield 2.48\% and 1.37\%, respectively (quite a bit better, especially when both skewness and excess kurtosis are adjusted for).

\begin{table}[t!]

\centering

\caption{\textit{Values of $n_3^*$ and $n_{ 34 }^*$ in the family income example, with four values of $\epsilon$ and three interesting quantiles.}}

\bigskip

\begin{tabular}{l||r|r||r|r||r|r}

& \multicolumn{2}{c||}{$z = \Phi^{ -1 } ( 0.975 )$} & \multicolumn{2}{c||}{$z = \Phi^{ -1 } ( 0.995 )$} & \multicolumn{2}{c}{$z = \Phi^{ -1 } ( 0.9995 )$} \\ \cline{2-7} 

\multicolumn{1}{c||}{$\epsilon$} & \multicolumn{1}{c|}{$n_3^*$} & \multicolumn{1}{c||}{$n_{ 34 }^*$} & \multicolumn{1}{c|}{$n_3^*$} & \multicolumn{1}{c||}{$n_{ 34 }^*$} & \multicolumn{1}{c|}{$n_3^*$} & \multicolumn{1}{c}{$n_{ 34 }^*$} \\

\hline

0.01 & 197 & 213 & 48 & 90 & 3 & 15 \\

0.005 & 788 & \textbf{821} & 190 & 279 & 9 & 36 \\

0.001 & 19,695 & 19,858 & 4,741 & \textbf{5,219} & 218 & 374 \\

0.0005 & 78,778 & 79,104 & 18,964 & 19,929 & 872 & \textbf{1,199}

\end{tabular}

\label{t:income-results-1}

\end{table}

\begin{figure}[t!]

\centering

\caption{\textit{Contour plot of $e^* ( n, z ) = | A_n ( z ) + B_n ( z )|$ on the CDF scale (equation (\ref{e:edgeworth-2})) in the income case study, when the skewness and excess kurtosis Edgeworth corrections are both made: the horizontal and vertical scales are $z$ and $\log_{ 10 } ( n )$, respectively.}}

\vspace*{-0.25in}

\includegraphics[ scale = 0.8 ]{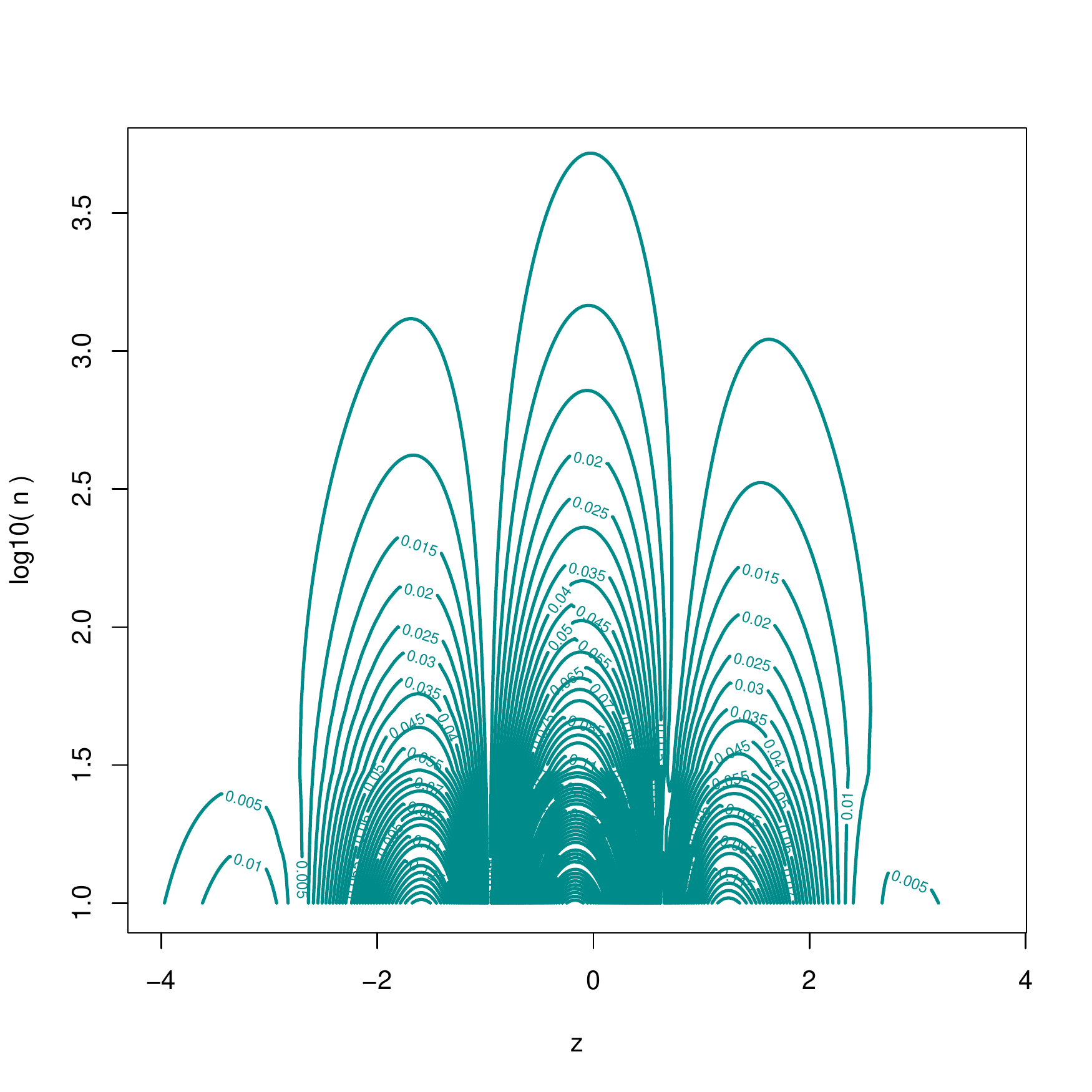}

\label{f:e-star-1}

\end{figure}

\begin{figure}[t!]

\centering

\caption{\textit{Perspective plot of the same $e^* ( n, z )$ as in Figure \ref{f:e-star-1}; the $e^*$ values range from near 0 (purple) to almost 0.3 (yellow).}}

\includegraphics[ scale = 0.7 ]{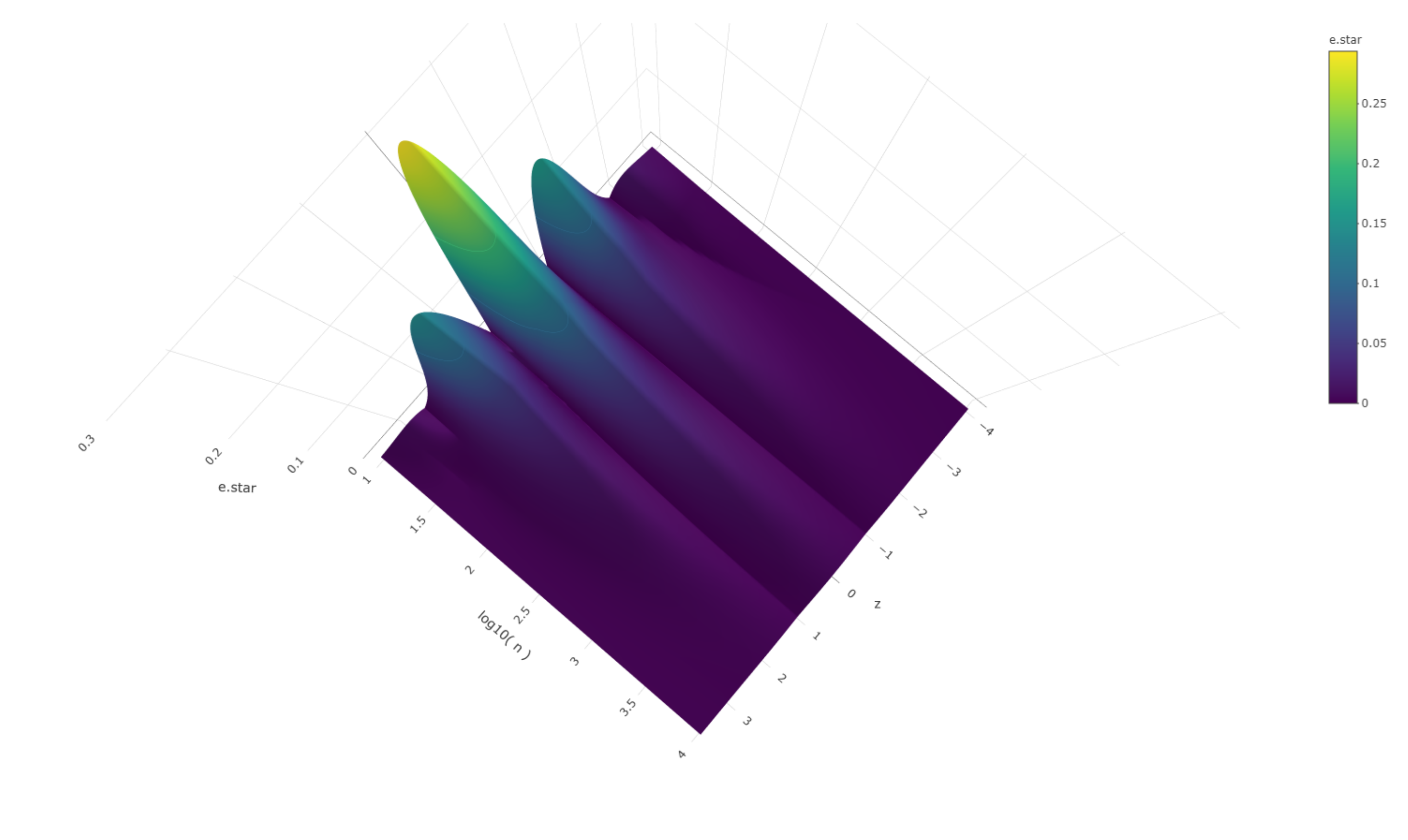}

\label{f:e-star-2}

\end{figure}

Referring back to the expressions in Section \ref{s:continuous-case}, how big does $n$ need to be so that the error $e^* ( n, z )$ on the CDF scale is at most $\epsilon$ with this population's $( \lambda, \eta )$ values? Table \ref{t:income-results-1} provides some numerical answers to this question, with $\epsilon = ( 0.01, 0.005, 0.001, 0.0005 )$ and the three quantiles corresponding to central intervals with probability content $( 0.95, 0.99, 0.999 )$; results are given for both $n_3^*$ and $n_{ 34 }^*$ for comparison, although (of course) the $n_{ 34 }^*$ columns are more accurate (note how little they differ from the corresponding $n_3^*$ columns; with this data set, skewness is more important as a departure from Normality than tail-weight). Figures \ref{f:e-star-1}--\ref{f:e-star-2} add to the story by presenting contour and perspective plots of $e^* ( n, z )$ in this case study (note the echo of Figure \ref{f:interesting-function-1} in these plots).

Table \ref{t:income-results-1} displays all of the monotonicities (as $\epsilon$ and $z$ change, and as one moves from $n_3^*$ to $n_{ 34 }^*$) to be expected from the findings in Section \ref{s:continuous-case}; in particular, dividing $\epsilon$ by 10, in moving from row 1 to row 3 and from row 2 to row 4, increases $n_3^*$ exactly (and $n_{ 34 }^*$ approximately) by a multiplicative factor of $10^2 = 100$. The most startling large entries in the Table do not provide useful advice on sample size determination (for instance, no one who is interested in getting the endpoints of the central 95\% interval right would set $\epsilon$ to 0.0005); we've highlighted in bold the entries most relevant to actual practice at the three quantiles addressed by the Table.

A natural first reaction to the small estimated sample sizes in Table \ref{t:income-results-1} is incredulity that much of anything can be learned about the right tail of such a highly-positively-skewed and heavy-tailed distribution with so little data. To explore the accuracy of the smallest entries in Table \ref{t:income-results-1}, we performed a simulation study, as follows:

\begin{itemize}

\item[(1)]

For each sample size value ($n$) from 1 to 50 (inclusive), we simulated $M =$ 10,000,000 means of IID samples of size $n$;

\item[(2)]

We then computed the empirical quantiles at the 0.9995 point of standardized (mean 0, SD 1) versions of each of the sample mean simulated distributions in step (1); and finally

\item[(3)]

We compared the empirical quantiles in step (2) with their skewness- and excess-kurtosis-corrected Cornish-Fisher approximations from Section \ref{s:clt-on-quantile-scale-1}.

\end{itemize}

\begin{figure}[t!]

\centering

\caption{\textit{Comparison of two estimates of the 0.9995 quantiles of the standardized sample mean distributions with samples of size $( n = 1, 2, \dots, 50 )$ from the income population. Solid green curve: skewness- and excess-kurtosis-corrected Cornish-Fisher approximation; red dotted curve: empirical estimates from simulation study; black horizontal dotted line: asymptotic quantile for (extremely) large $n$.}}

\vspace*{-0.25in}

\includegraphics[ scale = 0.6 ]{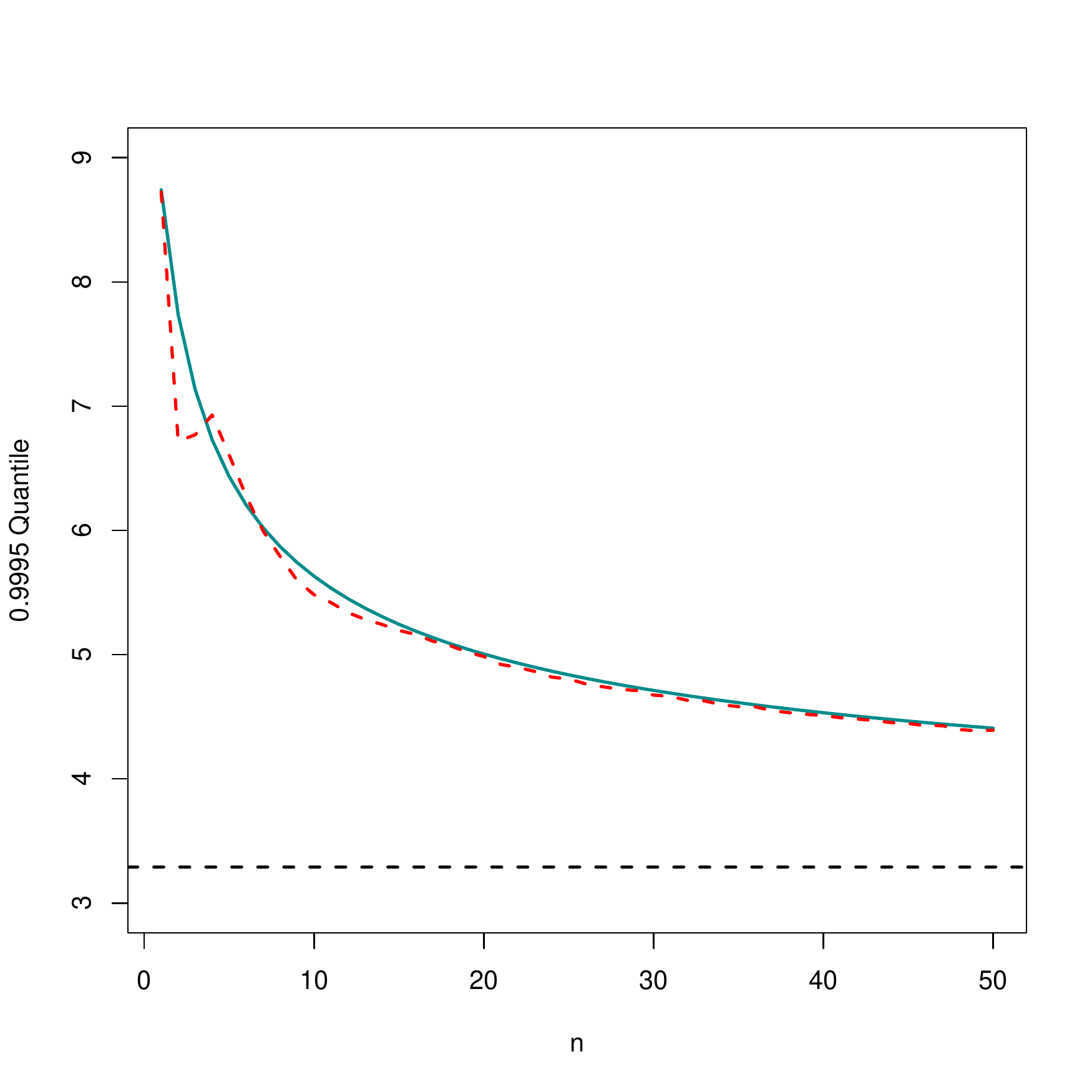}

\label{f:quantile-accuracy-1}

\end{figure}

Figure \ref{f:quantile-accuracy-1} summarizes the result of this simulation study. A comparison of the Cornish-Fisher approximations (solid green curve) with the empirical quantiles from the simulation (red dotted curve) shows that the process of convolving the sharply non-Normal income population PDF with itself $n$ times (the probabilistic operation underlying the calculation of $\bar{ Y }_n$) moves so aggressively toward the Normal distribution that the formulas in Section \ref{s:edgeworth-cornish-fisher} are already highly accurate in this case study with samples sizes as small as $n = 4$. 
 
% finish this: table 5 summarizes *probability* calculations, not
% statistical inference calculations: as soon as you have to
% *estimate* the population mean, sd, skewness, and excess kurtosis,
% everything is subject to O ( 1 / sqrt( n ) ) standard errors

\subsection{Roulette: red-or-black and single-number betting} \label{s:roulette}

In the game of \textit{roulette}, as implemented (for instance) in American casinos, a wheel with 38 painted and numbered slots --- 2 green (0, 00), 18 red (half of the integers from 1 to 36), and 18 black (the other half of those integers) --- into which a metal ball can fall is spun as a physical randomization device, in such a way that (when the wheel is well balanced) (a) all 38 possibilities are (to a good approximation) equally likely and (b) the spins of the wheel are logically, physically, and probabilistically independent. Gamblers (this is a game of pure chance, with no opportunities for skill) may bet on the outcome of the next spin before it occurs; there are a number of possible bets, all carefully arranged so that the gambler's expected net gain on any single wager of 1 monetary unit (MU) is $- \frac{ 2 }{ 38 }$, i.e., a net loss of about $-0.05226$ MU (the casino is also referred to as the \textit{house}, and $- \frac{ 2 }{ 38 }$ is called the \textit{house edge}; this value for roulette is remarkably low when compared with other wagering opportunities in casinos). If we keep track of the gambler's net gain on any single play with any of the available betting strategies, using random variables $( Y_i, i = 1, \dots, n )$, the PMF of the $Y_i$ will be a special case of the general class of 2--point discrete distributions (since the only possibilities are winning or losing), and this is the simplest environment in which to explore the lattice approximation results of Section \ref{s:discrete-case}.

Let the support of the $Y_i$ be $\{ v_1, v_2 \}$, in which the $v_j \ ( j = 1, 2 )$ are real numbers satisfying (without loss of generality) $v_1 < v_2$, and let $P ( Y_i = v_1 ) = ( 1 - p )$ and $P ( Y_i = v_2 ) = p$ for $0 < p < 1$; in the usual roulette situation, in which $v_1 < 0 < v_2$, $p$ represents the chance of the gambler coming out ahead on any single play. The discrete PMF of the $Y_i$ is clearly a lattice distribution (see \textbf{Definition \ref{d:lattice-distribution-1}}); the minimal lattice is specified by the choices $( a, h_{ max } ) = ( v_1, v_2 - v_1 )$. Set $S_n \triangleq \sum_{ i = 1 }^n Y_i$ and $\bar{ Y }_n \triangleq \frac{ S_n }{ n }$; $S_n$ keeps track of the gambler's cumulative net gain at the conclusion of $n$ plays.  

In this case study we focus on the gambler's chance of coming out ahead after $n$ plays, with two different types of bets: 
\begin{itemize}

\item

\textit{red-or-black}, in which a 1--MU wager on (say) red (\textit{R}) achieves a net gain of 1 MU with probability $p_R = \frac{ 18 }{ 38 }$ and suffers a net loss of 1 MU  with probability $( 1 - p_R )$, from which $( v_{ 1, R }, v_{ 2, R } ) = ( -1, +1 )$; this is the most conservative (risk-averse) betting strategy available in roulette; and

\item

\textit{single-number}, in which a 1--MU bet on any single number (\textit{SN}) from $\{ 0, 00, 1, \dots, 36 \}$ is rewarded with a net gain of $35$ MUs with probability $p_{ SN } = \frac{ 1 }{ 38 }$ and sustains a net loss of 1 MU with probability $( 1 - p_{ SN } )$, corresponding to $( v_{ 1, SN }, v_{ 2, SN } ) = ( -1, +35 )$; in sharp contrast with red-or-black, single-number wagering is the most aggressive (risk-seeking) roulette betting strategy.

\end{itemize}

\begin{figure}[t!]

\centering

\caption{\textit{Red-or-black betting in roulette with $n = 5$: broad-dotted black line, exact CDF of $Z_n$; narrow-dotted red curve, $O ( 1 )$ approximation; solid green line, $O \! \left( n^{ - \frac{ 1 }{ 2 } } \right)$ approximation from equation (\ref{e:lattice-correction-1}).}}

\vspace*{-0.15in}

\includegraphics[ scale = 0.8 ]{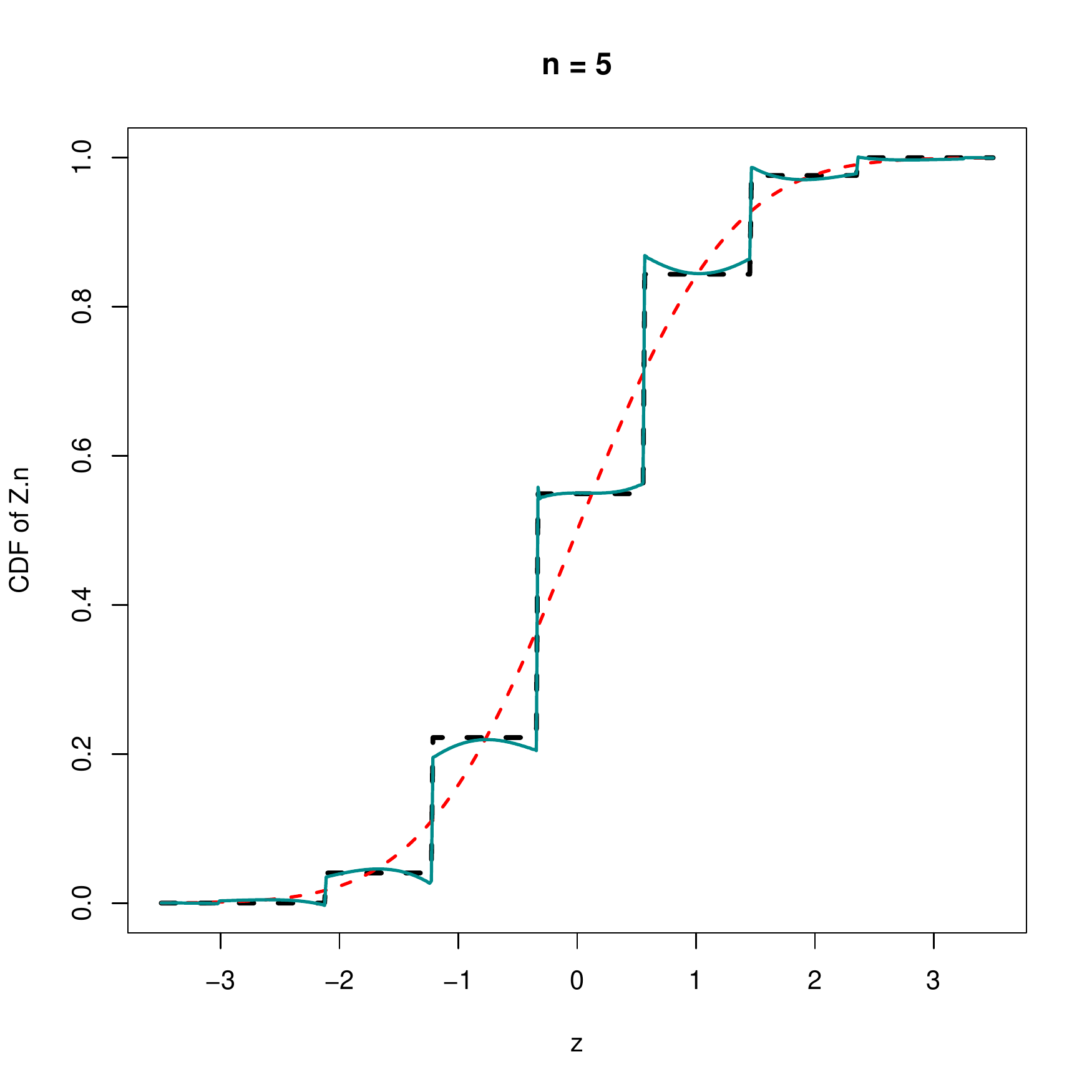}

\label{f:CDF-red-or-black-n-is-5-1}

\end{figure}

To use \textbf{Theorem \ref{t:edgeworth-lattice-1}} in this context, the $\bar{ Y }_n$ need to be standardized, so (as usual) set $Z_n = \frac{ \bar{ Y }_n - \mu }{ \sigma / \sqrt{ n } }$, in which $\mu = p \, ( v_2 - v_1 ) + v_1$ and $\sigma = ( v_2 - v_1 ) \, \sqrt{ p ( 1 - p ) }$ are the mean and SD of the $Y_i$, respectively; this determines 
\begin{equation} \label{e:a-star-and-h-max-star-1}
a^* = \frac{ a - \mu }{ \sigma } = - \frac{ p }{ \sqrt{ p \, ( 1 - p ) } } \ \ \ \textrm{and} \ \ \ h_{ max }^* = \frac{ h_{ max } }{ \sigma } = \frac{ 1 }{ \sqrt{ p \, ( 1 - p ) } } \, ,
\end{equation}
in which, interestingly, the standardization process has caused the $v_j$ to entirely vanish\footnote{Another way to think about what's going on is to define $Y_i^* \triangleq \frac{ Y_i - \mu }{ \sigma }$; the support of the $Y_i^*$ is the set $\{ v_1^*, v_2^* \} = \Big\{ - \frac{ p }{ \sqrt{ p \, ( 1 - p ) } }, \frac{ 1 - p  }{ \sqrt{ p \, ( 1 - p ) } } \Big\}$, from which (in parallel with the unstandardized values) $a^* = v_1^*$ and $h_{ max }^* = ( v_2^* - v_1^* )$.}. Figure \ref{f:CDF-red-or-black-n-is-5-1} presents the exact CDF of $Z_n$ with red-or-black betting and $n = 5$, together with two approximations: the $O ( 1 )$ result (the red dotted curve) obtained by naively using the CLT (\textbf{Theorem \ref{t:clt-on-cdf-scale-1}}) and the $O \! \left( n^{ - \frac{ 1 }{ 2 } } \right)$ approximation (the green solid curve) from \textbf{Theorem \ref{t:edgeworth-lattice-1}}. The latter approximation is remarkably good even with an $n$ of only 5, but note that this has been accomplished (for small $n$) by introducing two violations of basic CDF properties: (a) the $O \! \left( n^{ - \frac{ 1 }{ 2 } } \right)$ approximation should ideally be monotonic-non-decreasing and is not (note the slight convexity (for positive $z$) and concavity (when $z$ is negative) of the approximation in the intervals in which the exact CDF is flat), and (b) the $O \! \left( n^{ - \frac{ 1 }{ 2 } } \right)$ approximation goes slightly below 0 and slightly above 1 at the edges of the plot (these defects of course disappear with larger $n$).

\begin{figure}[t!]

\centering

\caption{\textit{Single-number betting in roulette with $n = 5$ (top panel) and $n = 100$ (bottom panel): broad dotted black line, exact CDF of $Z_n$; narrow dotted red line, $O ( 1 )$ approximation; solid green line, $O \! \left( n^{ - \frac{ 1 }{ 2 } } \right)$ approximation from equation (\ref{e:lattice-correction-1}).}}

\vspace*{-0.15in}

\includegraphics[ scale = 0.8 ]{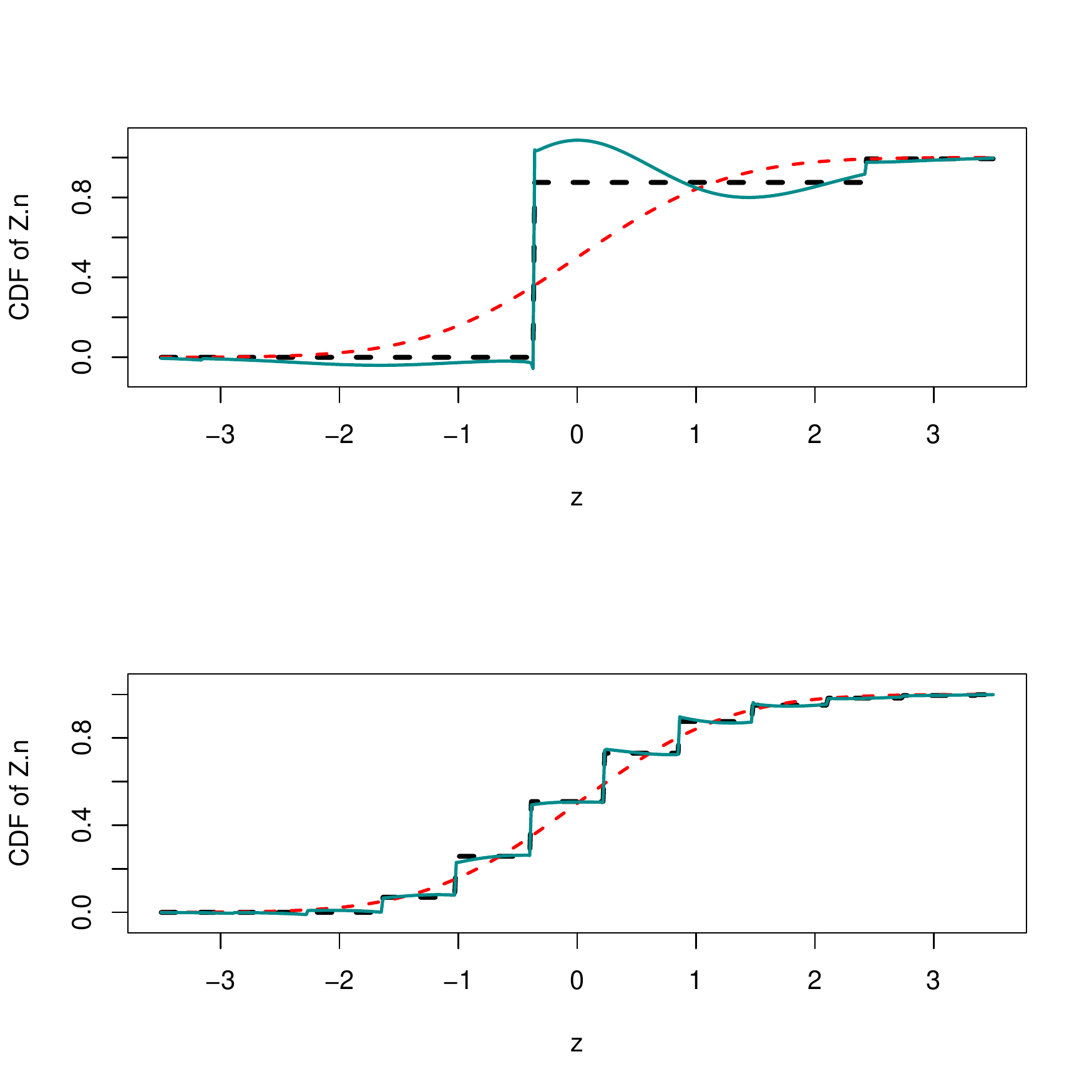}

\label{f:CDF-single-number-n-is-5-and-100-1}

\end{figure}

Figure \ref{f:CDF-single-number-n-is-5-and-100-1} presents results analogous to those in Figure \ref{f:CDF-red-or-black-n-is-5-1}, but this time examining single-number betting with $n = 5$ (top panel) and $n = 100$ (bottom panel). Because the maximal span for single-number wagering on the MU scale is $\frac{ 35 - ( -1 ) }{ 1 - ( -1 ) } = 18$ times larger than for red-or-black --- or, equivalently, because $p_{ SN }$ is so much farther away from $\frac{ 1 }{ 2 }$ than $p_{ R }$ --- the lattice result is substantially less accurate with $n = 5$ in this new plot: as with red-or-black wagering, in addition to violating non-decreasing monotonicity, the $O \! \left( n^{ - \frac{ 1 }{ 2 } } \right)$ approximation goes negative in a small region of negative $z$ and exceeds 1 in a small $z$ interval for $z > 0$, but these violations are noticeably more extreme with single-number betting. The behavior of this approximation between the two jump points in the exact CDF in the upper panel of the Figure nicely illustrates the Fourier-series representation of the $J ( \cdot )$ function in Section \ref{s:discrete-case}. You can see from the bottom panel in Figure \ref{f:CDF-single-number-n-is-5-and-100-1} that, by the time $n$ has reached 100 in single-number betting, the $O \! \left( n^{ - \frac{ 1 }{ 2 } } \right)$ approximation has attained about the same accuracy as with $n = 5$ in red-or-black.

Our principal interest in this case study is the exact and approximate computation of
\begin{equation} \label{e:the-point}
\theta_n \triangleq P ( \textrm{gambler comes out ahead after } n \textrm{ plays} ) = P ( S_n > 0 ) = P ( \bar{ Y }_n > 0 ) \, .
\end{equation}
Note that the affine transformation 
\begin{equation} \label{e:bernoulli-2}
X_i \triangleq \frac{ Y_i - a }{ h_{ max } } = \frac{ Y_i - v_1 }{ v_2 - v_1 } \ \ \ \textrm{yields} \ \ \ X_i \stackrel{ \textrm{\scriptsize IID} }{ \sim } \textrm{Bernoulli} ( p ) \, ;
\end{equation} 
this makes it possible to obtain an exact expression for the desired probability in (\ref{e:the-point}) in terms of the CDF of the Binomial distribution\footnote{This in turn involves evaluating an Incomplete Beta function, which has no closed-form analytic expression, but approximations of Binomial tail probabilities are available to essentially any desired practically-relevant precision in standard software packages (e.g., the \texttt{pbinom} function in \texttt{R}).}). Set $T_n \triangleq \sum_{ i = 1 }^n X_i$, noting that $T_n \sim \textrm{Binomial} ( n, p )$; then, inverting the affine transformation in equation (\ref{e:bernoulli-2}),
\begin{eqnarray} \label{e:bernoulli-3}
\theta_n = P ( S_n > 0 ) & = & P \left\{ \sum_{ i = 1 }^n \left( a + h_{ max } \, X_i \right) > 0 \right\}\nonumber \\
& = & P \left( T_n > - \frac{ a \, n }{ h_{ max } } \right) = 1 - F_{ T_n } \left( - \frac{ v_1 \, n }{ v_2 - v_1 } \right) \, ,
\end{eqnarray} 
in which $F_{ T_n } ( t )$ is the CDF of the Binomial$( n, p )$ distribution.

\begin{figure}[t!]

\centering

\caption{\textit{$\theta_n =$ P(coming out ahead after $n$ plays) (exact value, solid red curve; $O ( 1 )$ approximation, solid green curve) as a function of $n$: left panel, red-or-black ($n$ from 1 to 50); right panel, single-number ($n$ from 1 to 500). \textbf{NB} The horizontal scales in the left and right panels are sharply different.}}

\bigskip

\includegraphics[ scale = 0.8 ]{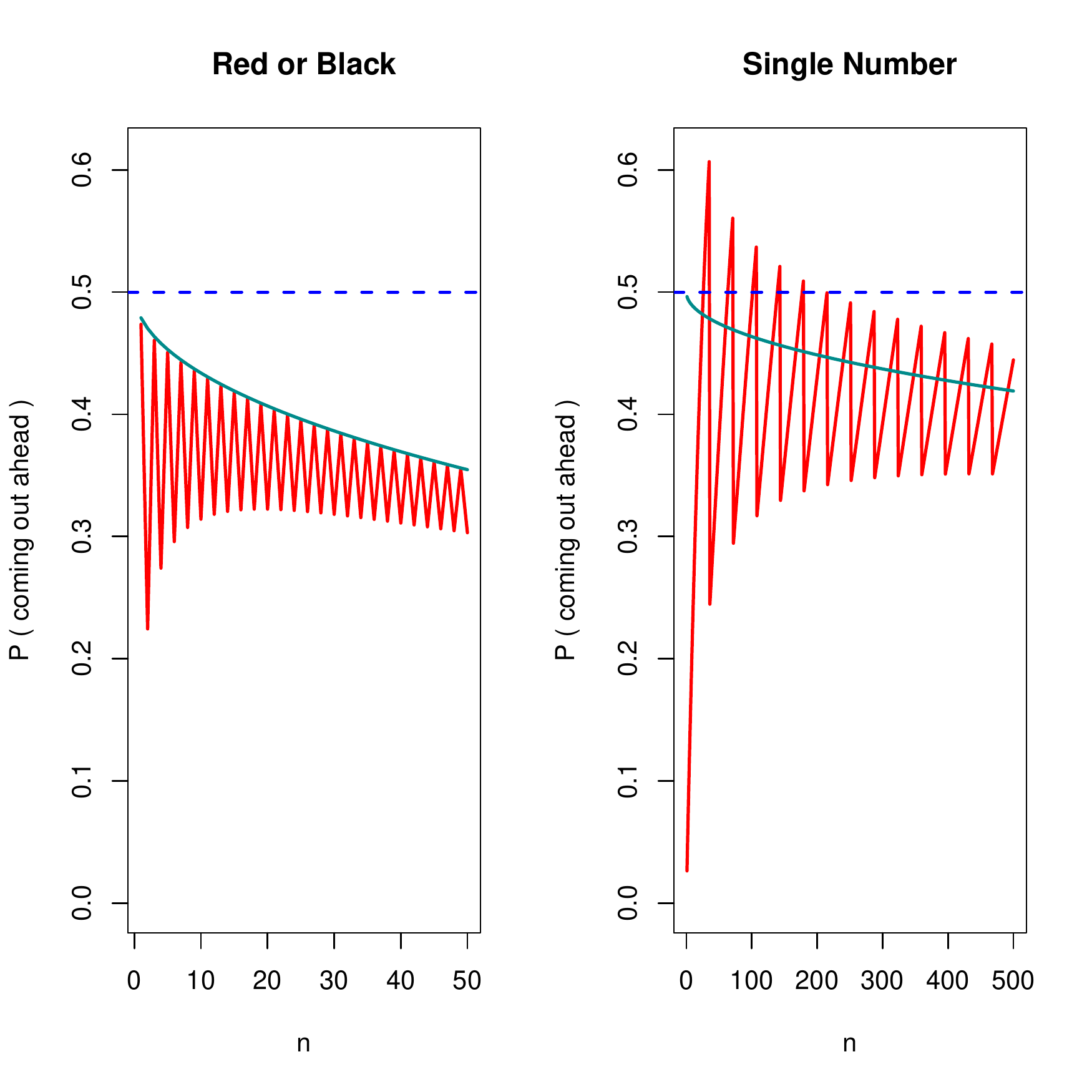}

\label{f:coming-out-ahead-1}

\end{figure}

Figure \ref{f:coming-out-ahead-1} plots $\theta_n$ against $n$, with red-or-black and single-number gambles in the left- and right-hand panels, respectively. If you haven't seen this plot before, it's rather startling, in four ways:
\begin{itemize}

\item

Note the damped-oscillation behavior of $\theta_n$ as $n$ increases, which is a direct consequence of the lattice character of the PMF of $S_n$; for each of the two gambles illustrated, the period of the oscillation is driven partly by (a) the span of the lattice and (b) the success probability on each play;

\item

Note further the remarkable, and at first sight too-good-to-be-true, result with single-number betting that there are quite a few small values of $n$ for which the gambler's chance of coming out ahead exceeds 0.5; indeed the largest of these values is 0.607 and occurs at $n = 35$ (the fact that this equals $v_{ 2, SN }$ is not a coincidence). The explanation for this apparent paradox is that the casino is perfectly happy to offer you betting strategies with $\theta_n > 0.5$, because all that the house cares about is your expected net gain, which is comfortably (from the casino's point of view) negative for all possible bets in roulette. Here, for example, using the $- \frac{ 2 }{ 38 }$ fact mentioned above as the gambler's expected net gain on any single 1--MU play, with $n = 35$ the casino \textit{expects} (i.e., on average) to make $\frac{ 2 \cdot 35 }{ 38 } \doteq 1.84$ MUs from any gambler attempting the single-number strategy\footnote{Calculation reveals that the single most likely outcome with this strategy is that you'll lose 35 MUs (with probability 0.394), and that --- if you do win --- by far the most likely amount you'll win is 1 MU (with (unconditional) probability 0.372).} with $n = 35$;

\item

Note the behavior of $\theta_n$ as $n$ increases: since every bet has an expected net loss, $\lim_{ n \rightarrow \infty } \theta_n = 0$ for all betting strategies in roulette that involve making the same bet over and over\footnote{The stochastic process $\{ S_n, n = 1, 2, \dots \}$ that keeps track of the gambler's net gain at the end of $n$ plays is a random walk and is therefore first-order Markovian; this provides another approach (not pursued here) to thinking about $\theta_n$.}, but the \textit{rate} at which this limiting behavior is achieved is amazingly slow. For instance, with red-or-black wagering, you still have a 30.0\% chance of coming out ahead after 99 plays, and this probability is still more than 12\% with $n = 495$. The corresponding values for single-number betting\footnote{This is an example of a \textit{bold-play-is-optimal} result of the type investigated by \cite{dubins-savage-1965}: if all of the gambles available to you have the same negative expected net loss per play, you will maximize your chance of coming out ahead by finding the gamble with the biggest SD($Y_i$) (earlier this was referred to as risk-seeking behavior). Of course, by approximate symmetry of the PMF of $S_n$, this will also maximize your chance of losing a bundle.} are even more surprising: $n = [ 1,000 ; 5,000 ; 10,000 ] \leftrightarrow \theta_n = ( 0.396, 0.268, 0.185 )$; and

\item

Finally, imagine creating two curves, one in each panel of Figure \ref{f:coming-out-ahead-1}, that pass through the midpoints of the oscillations as you scan from left to right (i.e., as $n$ increases); this process yields a kind of locally smoothed estimate of $\theta_n$ as a function of $n$. Compare those curves in each panel with the $O ( 1 )$ approximations (the green functions); for both betting strategies, it's fair to say informally that the $O ( 1 )$ results are ``biased on the high side.'' It's intuitively surprising that the positive bias is much larger for red-or-black betting (in which the chance of coming out ahead on any single play is so close to $\frac{ 1 }{ 2 }$) than for single-number gambles.

\end{itemize}

\begin{figure}[t!]

\centering

\caption{\textit{$\theta_n ( \epsilon ) =$ P(net gain of at least $\epsilon$ after $n$ plays) as a function of $n$ for $n = ( 1, \dots, 20 )$ with red-or-black betting; in all three panels, the exact value is plotted as a solid red curve, the $O ( 1 )$ approximation is the solid green curve, the approximation out to order $O \! \left( n^{ - \frac{ 1 }{ 2 } } \right)$ but including only the skewness correction is the purple dotted curve, and the full \x \ approximation with both skewness and lattice corrections is the black dotted curve. Top panel: $\epsilon = 0$; middle panel: $\epsilon = 0.01$; bottom panel: $\epsilon = 0.5$.}}

\includegraphics[ scale = 0.8 ]{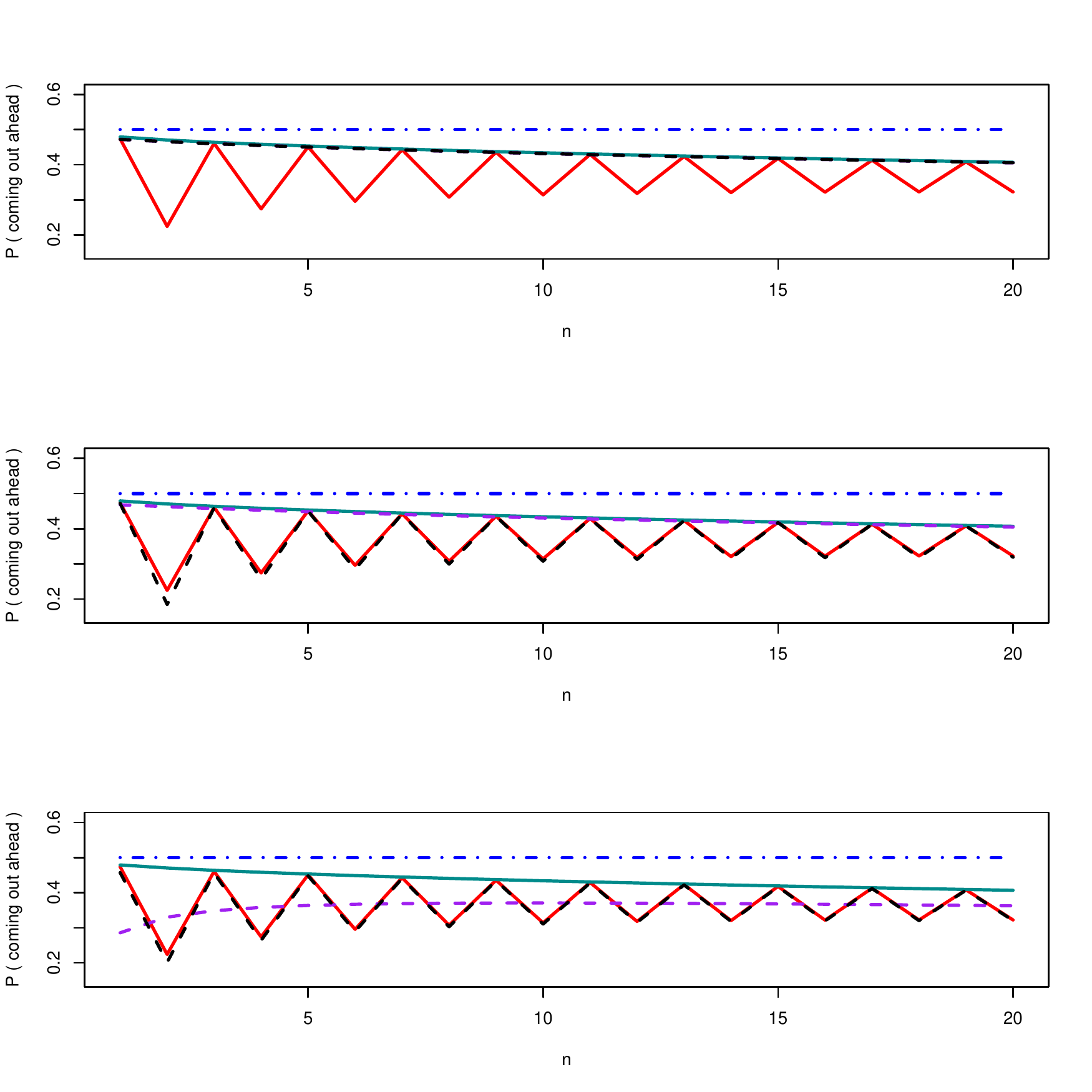}

\label{f:coming-out-ahead-2}

\end{figure}

Adding more accurate approximations to Figure \ref{f:coming-out-ahead-1} yields another layer of complexity: for $0 \le \epsilon < 1$, Figure \ref{f:coming-out-ahead-2} generalizes $\theta_n$ to $\theta_n ( \epsilon ) \triangleq P$(net gain of at least $\epsilon$ after $n$ plays) and examines the effect of $\epsilon$ with red-or-black wagering. In the top panel, in which $\epsilon = 0$, the skewness correction is tiny, which leads to contradictory intuitions: on the one hand, this makes sense, since (as noted above) $p_{ R }$ is so close to $\frac{ 1 }{ 2 }$, but --- on the other hand --- the left panel of Figure \ref{f:coming-out-ahead-1} makes it abundantly clear that this is precisely a situation in which a large skewness correction to the $O ( 1 )$ approximation is needed. Furthermore, this panel also documents the initially surprising fact that, for this value of $\epsilon$, the lattice correction has \textit{absolutely no effect} on the \x \ approximation to $\theta_n ( 0 )$. The technical reason is as follows:
\begin{equation} \label{e:weird-result-1}
1 - F_{ Z_n } ( z ) = P \left(\frac{ \bar{ Y }_n - \mu }{ \sigma / \sqrt{ n } } > z \right) = P (S_n > \sigma \, z \, \sqrt{ n } + n \, \mu ) \, .
\end{equation}
Now $\sigma \, z \, \sqrt{ n } + n \, \mu = 0$ iff $z \triangleq z^* = - \frac{ \mu \, \sqrt{ n } }{ \sigma }$, and this is the $z$ value that enters the expression $J \left( \frac{ z^* \, \sqrt{ n } - n \, a^* }{ h_{ max }^* } \right)$ in equation (\ref{e:lattice-correction-1}) in computing the lattice correction; substituting in the formulas (defined above) for $( \mu, \sigma, a^*, h_{ max }^*, z^* )$ in the special case of red-or-black betting and simplifying, the result is that the contribution to the lattice correction from the $J$ function is simply $J \! \left( \frac{ n }{ 2 } \right)$, which is 0 for all $n$.

The middle and bottom panels of Figure \ref{f:coming-out-ahead-2} document what happens to $\theta_n ( \epsilon )$ for $\epsilon = 0.01$ (middle panel) and 0.50 (bottom panel). Simply increasing $\epsilon$ from 0 to 0.01 causes the lattice approximation to appear, and to be remarkably accurate even for quite small values of $n$ (although the $O ( 1 )$ approximation with skewness correction is still noticeably biased (in the sense defined earlier). The bottom panel illustrates the disappearance of the bias when $\epsilon$ move to 0.5, and the combination of the two \x \ corrections is now almost perfect.

\begin{figure}[t!]

\centering

\caption{\textit{$\theta_n = P$(coming out ahead after $n$ plays) as a function of $n$, for $n = ( 1, \dots, 200 )$ with single-number wagering; as in the previous Figure, the exact value is plotted as a solid red curve, the $O ( 1 )$ approximation is the solid green curve, the approximation out to order $O \! \left( n^{ - \frac{ 1 }{ 2 } } \right)$ but including only the skewness correction is the purple dotted curve, and the full \x \ approximation with both skewness and lattice corrections is the black dotted curve. Vertical dotted orange lines are plotted at points of the form $n = k \left(  \frac{ h_{ max } }{ 2 } \right)$ for $k$ a positive integer.}}

\vspace*{-0.2in}

\includegraphics[ scale = 0.8 ]{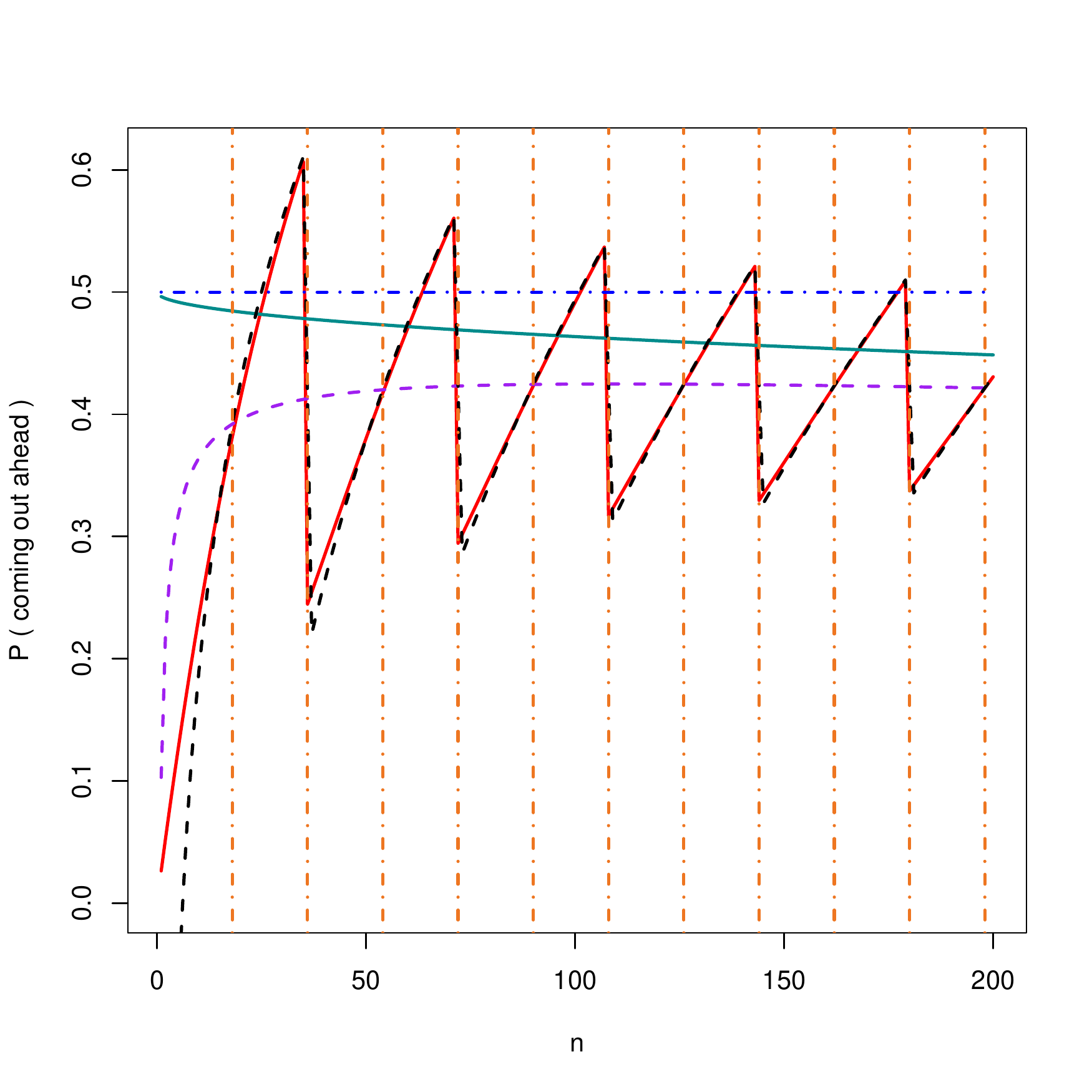}

\label{f:coming-out-ahead-3}

\end{figure}

Figure \ref{f:coming-out-ahead-3} summarizes the behavior of the approximations to $\theta_n$ ($\epsilon = 0$) for single-number betting (the unusual behavior of the lattice correction in red-or-black wagering when $S_n$ is compared with 0 does not appear here, because 0 is no longer the midpoint of the $( v_1, v_2 )$ interval). As usual, the $O ( 1 )$ approximation (the solid green curve) is biased high; this bias is absent with this betting strategy (see the purple dotted curve in the Figure) when the skewness correction is applied, except for quite small values of $n$. The full \x \ approximation is remarkably accurate for $n \ge 15$.

%In the Appendix we detail several pitfalls that we encountered in implementing the lattice-distribution theoretical results of Section \ref{s:discrete-case} in the roulette case study, in the hope that others may avoid these difficulties.

% several?

\section{Discussion} \label{s:discussion}

The revolution in thought begun with Jacob Bernoulli's realization around the year 1700 that stable relative frequencies in IID gambling settings could be generalized to problems in ``civil, moral and economic affairs'' led over the subsequent 300 years to an increasingly sophisticated mathematical analysis of the \textit{rate} at which sample means approach population means, culminating in the versions of the Central Limit Theorem (CLT) in vigorous use today. In two real-world-relevant case studies 
we've documented the extent to which Edgeworth and Cornish-Fisher adjustments to $O ( 1 )$ CLT approximations have proven themselves to be remarkably accurate in practical data science probability calculations involving both discrete and continuous sampling distributions, even with quite small sample sizes. We encourage other researchers to broaden and deepen the scope of our empirical findings, for example by examining case studies involving more complicated stochastic processes than IID (e.g., CLTs for processes with Markovian, multivariate, strong-mixing, or martingale characters and for processes unfolding on generalizations of $\mathbb{ R }^k$ such as manifolds).

\theendnotes

\section*{Acknowledgments}

We have frequently relied in the history section on the work of Anders Hald; we thank him for his careful and interesting scholarship. We're grateful to John Kolassa, and Robert Lund for helpful references and comments. Membership on this list does not constitute agreement with the views presented here, nor are any of these people responsible for any errors that may remain.

% fix the capitalizations in the .bib file

\bibliography{practical-scope-of-clt}

\end{document}